\documentclass[12pt]{article}
\usepackage{graphicx}
\usepackage{amsmath}
\usepackage{amssymb}
\usepackage[numbers,compress]{natbib}
\usepackage{amsmath}
\begin{document}
\title{Time evolution of the free Dirac field in spatially flat FLRW space-times}

\author{Ion I. Cot\u aescu \footnote{e-mail: i.cotaescu@e-uvt.ro}\\
{\small \it West University of Timi\c soara,}\\{\small \it V. P\^ arvan Ave. 4, RO-300223, Timi\c soara, Romania}}
%

\maketitle

\begin{abstract}
The framework of the relativistic quantum mechanics on spatially flat FLRW space-times is considered for deriving  the analytical solutions of the Dirac equation in different local charts of these manifolds.  Systems of commuting conserved operators are used for determining the fundamental solutions as common eigenspinors giving thus physical meaning to the integration constants related to the eigenvalues of these operators. Since these systems, in general, are incomplete on the FLRW space-times there are integration constants that must be fixed by setting the vacuum either as the traditional adiabatic one or as the rest frame vacuum we proposed recently. All the known solutions of the Dirac equation on these manifolds are discussed in all details and a new type of spherical waves of given energy in the de Sitter expanding universe is reported here for the first time.  
 
Pacs: 04.62.+v
\end{abstract}

\newpage

\tableofcontents

\newpage
\section{Introduction}

In general relativity the Dirac field on Riemannian manifolds is less studied   \cite{nach,OT,BADU,ALG,SH,Vil,SH1,FGV,KT} since one prefers the scalar field  that can be manipulated easier in the actual developments in astrophysics and cosmology. For this reason we devote this paper to the Dirac field on the spatially flat Friedmann-Lema\^ itre-Robertson-Walker (FLRW) space-times paying a special attention to the framework we need for studying the analytical solutions of the Dirac equations that can be derived on such manifolds \cite{CD1,CD2,CD3,CD4,CD5,CD6,CD7},  including a new type of solution reported here for the first time.

In our opinion, the general relativistic quantum mechanics on curved space-times must respect {\em ad litteram} the principles of the traditional quantum mechanics which today is more than a coherent theory being in fact the source of new technologies. On the other hand, an independent relativistic quantum mechanics cannot be constructed since the wave functions are replaced here by free fields with different spins and specific equations. For this reason, the general relativistic quantum mechanics must be seen as the one particle restriction of the quantum field theory (QFT) at the level of the first quantization. This theory must be compatible with the geometry of any curved background, keeping a balance between the local and global features.

We assume that the quantum states are prepared or measured by a global apparatus represented by algebra of the conserved operators (or observables) commuting with the operator of the field equation.  The principal  observables are the  differential operators  generated by the Killing vector fields  but, in addition,  there are observables that can be defined in different manners as we shall see in what follows. All these observables must be defined globally, independent on the local charts we use. The solutions of the field equations have to be determined completely or partially as common eigenstates of a system of commuting observables since then the integration constants get a physical meaning as eigenvalues of these operators. Thus we obtain complete systems of fundamental solutions, globally defined,  representing the bases the different representations (reps.) of the quantum mechanics and QFT.     

On the other hand,  the time evolution can be described in many time evolution pictures which are strongly dependent on the local time coordinate. For this reason we assume that two local charts may generate different time evolution pictures if their coordinates are related through a diffeomorphism transforming simultaneously the time and space coordinates.  Thus we are in the apparently paradoxical situation to work in global reps. but with local time evolution pictures. In what follows we would like to investigate exhaustively the solutions of the Dirac equation on the spatially flat FLRW  manifolds, including the de Sitter (dS) one,  in two different time evolution pictures, namely the natural picture (NP)  in co-moving charts and the Sch\" rodinger picture (SP) \cite{CD3} in the charts with de Sitter-Painlev\' e (dS-P)  coordinates \cite{dSP1,dSP2}. We revisit the known plane \cite{nach,CD1} and spherical \cite{SH,CD2} waves depending on vector or scalar momentum, $P$,  in NP and the plane waves depending on energy, $E$,  in the SP of the dS expanding universe \cite{CD4}. Moreover, on this manifold we derive for the first time new spherical waves depending on energy in the SP where the spherical variables can be separated. For shortening the terminology we shall speak about plane and spherical P-waves in NP and plane and spherical E-waves in SP or NP of the dS background. These sets of fundamental solutions define  different reps. among them  those of  plane P-waves and E-waves can be related in NP while for the spherical solutions we cannot obtain a similar result.  

On the manifolds into consideration here we find only  incomplete sets of commuting observables such that there remain undetermined integration constants which must be fixed  by using supplemental assumptions  related mainly to the frequencies separation setting the vacuum. In what follows we consider the traditional adiabatic vacuum (a.v.) \cite{BuD,BD} and the new rest frame vacuum (r.f.v.) we proposed recently \cite{CrfvD}. By using these vacua we can determine the integration constants of all the plane waves but there are some ambiguities in the case of the spherical waves we will discuss here. 

We start in next section with the geometry of the spatially flat FLRW space-times defining the frames we need for writing the gauge covariant Dirac field \cite{ES} whose conserved observables are briefly analysed showing how these may define the reps. of our relativistic quantum mechanics. In section 3 we introduce the time evolution pictures, NP and SP,  we need in order to avoid coordinate transformations involving time. In this section we define, in addition, the energy operators and study the equivalence of these pictures. The next section is devoted to the solutions that can be obtained in the NP presenting plane and spheric P-waves  for which we define the a.v. and r.f.v. giving as examples a Milne-type manifold and the dS expanding universe.  The plane and spherical E-waves, that can be derived exclusively  in the SP of this last manifold, are presented in section 5, including the new solutions derived here.  We point out that these depend on the conserved energy which separates the frequencies as in special relativity  but leaving undetermined an integration constant. Finally we present our concluding remarks and present some technical details in four appendices.


\section{Dirac field in FLRW space-times}

We study here the free Dirac field (or perturbation) on the $(1+3)$-dimensional local Minkowskian spatially flat FLRW space-times whose geometries are given by a scale factor $a: D_t \to {\Bbb R}$ which is a smooth function defined on a given time domain $D_t$. We denote from now these FLRW space-times by $(M, a)$ for distinguish them from the general Riemannian manifolds of arbitrary metric $g$ denoted by $(M,g)$. The Minkowski space-time will be denote by $(M,\eta)$ adopting the following signature $\eta={\rm diag}(1,-1,-1,-1)$ for its metric.

\subsection{Frames in FLRW space-times}

The form of the Dirac equation  depends on the choice of the  the local coordinates and the unholonomic orthogonal local frames needed for describing the spin. On the FLRW manifolds $(M,a)$ there are many types of local charts (or natural frames) related to the standard co-moving FLRW one, $\{t,{\bf x}\}$, whose coordinates $x^{\kappa}$ (labelled  by the natural indices $\kappa,\nu,...=0,1,2,3 $) are the proper (or cosmic) time, $t$, and the conformal Cartesian space coordinates $x^i$  ($i,j,k...=1,2,3$)  for which we  use the vector notation ${\bf x}=(x^1,x^2,x^3)$. Another useful chart is that of the conformal time,
\begin{equation}\label{tct}
t_c  =\int \frac{dt}{a(t)}~\to~  a(t_c  )=a[t(t_c  )]\,,
\end{equation}
and the same Cartesian space coordinates, denoted by $\{t_c  ,{\bf x}\}$. The line elements of these charts are,  
\begin{eqnarray}
ds^2=g_{\kappa\nu}(x)dx^{\kappa}dx^{\nu}&=&dt^2-a(t)^2 d{\bf x}\cdot d{\bf x}\label{dsxt}\\
&=&a(t_c)^2\,(dt_c  ^2-d{\bf x}\cdot d{\bf x})\,.\label{dsxtc}
\end{eqnarray}
The advantage of the conformal chart is that here one can take over many results from the flat Minkowski space-time $(M,\eta)$ through a simple conformal transformation \cite{BD}.

A less used chart is the dS-P one, $\{t, \underline{\bf x}\}$,  with 'observed' space coordinates defined as \cite{dSP1},
\begin{equation}\label{xx}
\underline{x}^i=a(t) x^i\,. 
\end{equation}
When we change the coordinates $\{t, {\bf x}\}\to \{t, \underline{\bf x}\}$ the time remains the same such that
\begin{equation}\label{X1}
\underline{\partial}_i\equiv \frac{\partial}{\partial \underline{x}^i}=\frac{1}{a(t)}\partial_i\,.
\end{equation}
The  line element
\begin{equation}\label{dsxot}
ds^2=dt^2 \left(1-\frac {\dot{a}(t)^2}{a(t)^2}\, \underline{\bf x}\cdot{\bf \underline{x}}\right)+2\,\frac{\dot{a}(t)}{a(t)}\, \underline{\bf x}\cdot d \underline{\bf x} \,dt-d \underline{\bf x}\cdot d \underline{\bf x}\,,
\end{equation}
where $\dot{a}(t)=\partial_t a(t)$, depends on Hubble function $\frac{\dot{a}(t)}{a(t)}$   for which we do not use the symbol $H$ since this is reserved for the energy operators. The corresponding chart  $\{t_c  , \underline{\bf x}\}$, with the same space coordinates but with the conformal time, has the line element 
\begin{equation}\label{dsxotc} 
ds^2=dt_c^2 \left(a(t_c)^2-\frac{\dot{a}(t_c)^2}{a(t_c)^2}\, \underline{\bf x}\cdot  \underline{\bf x}\right)+2\,\frac{\dot{a}(t_c)}{a(t_c)}\, \underline{\bf x}\cdot d{\bf \underline{x}} \,dt_c-d \underline{\bf x}\cdot d \underline{\bf x}\,,
\end{equation}
since after changing the time variable $t\to t_c  $ and denoting $\dot{a}(t_c  )=\partial_{t_c  }a(t_c )$,  we have to substitute  $dt \to a(t_c) dt_c $ and
\begin{equation}\label{subs}
\partial_t\to \frac{1}{a(t_c)}\,\partial_{t_c}\,,\quad \frac{\dot{a}(t)}{a(t)}\to \frac{\dot{a}(t_c  )}{a(t_c  )^2}\,,
\end{equation} 
in accordance with Eq. (\ref{tct}).

For studying problems with spherical symmetry we need local charts with spherical coordinates,  $\{t,r,\theta, \phi\}$ and $\{t_c,r,\theta, \phi\}$,  obtained from  the charts $\{t,{\bf x}\}$ and $\{t_c,{\bf x}\}$ where we introduce the spherical coordinates ${\bf x}\to (r,\theta,\phi)$ with $r=|{\bf x}|$. In the chars with dS-P coordinates, $\{t,\underline{\bf x}\}$ and $\{t_c,\underline{\bf x}\}$, we consider the coordinates $\underline{\bf x}\to (\underline{r},\theta,\phi)$ with $\underline{r}=|\underline{\bf x}|$ but the same angular coordinates. 

For writing down the Dirac equation  we need to fix the tetrad gauge giving  the vector fields $e_{\hat\alpha}=e_{\hat\alpha}^{\kappa}\partial_{\kappa}$ defining the local orthogonal frames,  and the 1-forms $\omega^{\hat\alpha}=\hat e_{\kappa}^{\hat\alpha}dx^{\kappa}$ of the dual co-frames (labelled by the local indices, $\hat\kappa,\hat\nu,...=0,1,2,3$). The metric tensor of $(M,g)$ can be expressed now as $g_{\kappa\nu}=\eta_{\hat\alpha\hat\beta}\hat e^{\hat\alpha}_{\kappa}\hat e^{\hat\beta}_{\nu}$.   Here we restrict ourselves to the diagonal tetrad gauge defined by the vector fields  
\begin{eqnarray}
e_0&=&\partial_t\,,\nonumber \\
e_i&=&\frac{\textstyle 1}{\textstyle a(t)}\,\partial_i=\underline{\partial}_{\,i}+\frac{\textstyle \dot{a}(t)}{\textstyle a(t)}\underline{x}^i\partial_t\,,\label{e}
\end{eqnarray}
and the corresponding dual 1-forms 
\begin{eqnarray}
\tilde\omega^0 &=&dt\,,\nonumber \\
\tilde \omega^i &=& a(t)dx^i=d\underline{x}^i-\frac{\dot{a}(t)}{a(t)}\underline{x}^i dt\,,\label{o}
\end{eqnarray}
in order to preserve the global $SO(3)$ symmetry  allowing us to use systematically the  $SO(3)$ vectors.  When we use the chars with spherical coordinates we keep the same gauge rewritten in terms of these coordinates.

The  frames  $\{x;e\}$ we need in the Dirac theory are formed by a local chart $\{x\}$ and the local frames defined by the tetrads $e$. These frames can be transformed, $\{x;e\}\to \{x';e'\}$, with the help of  diffeomorphis, $x\to x'=\phi(x)$, transforming the coordinates and by using  local transformations $\Lambda(x)$ of the Lorentz group, $L^{\uparrow}_{+}$, for changing the tetrad gauge as 
\begin{eqnarray}
e_{\hat\alpha}(x)&\to & e_{\hat\alpha}'(x)=\Lambda_{\hat\alpha\,\cdot}^{\cdot\,\hat\beta}(x)\, e_{\hat\beta}(x)\,,\label{gauge1}\\
{\tilde\omega}^{\hat\alpha}(x)&\to &{\tilde \omega}^{\prime\,\hat\alpha}(x)=\Lambda^{\hat\alpha\,\cdot}_{\cdot\,\hat\beta}(x) \,\tilde\omega_{\hat\beta}(x)\,.\label{gauge2}
\end{eqnarray}
In general, any theory of fields with spin in general relativity must be gauge invariant in the sense that the above gauge transformations have to do not affect the physical meaning of the theory.

\subsection{Covariant Dirac field}

In any frame $\{x;e\}$ of a curved space-time $(M,g)$ the tetrad gauge invariant action of the Dirac field $\psi$, of mass $m$, minimally coupled to the background gravity, reads,  
\begin{equation}\label{action}
{\cal S}[e,\psi]=\int\, d^{4}x\sqrt{g}\left\{
\frac{i}{2}[\overline{\psi}\gamma^{\hat\alpha}\nabla_{\hat\alpha}\psi-
(\overline{\nabla_{\hat\alpha}\psi})\gamma^{\hat\alpha}\psi] -
m\overline{\psi}\psi\right\}
\end{equation}
where  $\bar{\psi}=\psi^+\gamma^0$ is the Dirac adjoint of $\psi$ and  $g=|\det(g_{\mu\nu})|$. The  Dirac matrices, $\gamma^{\hat\alpha}$ (with local indices) are self adjoint, $\overline{\gamma^{\hat\mu}}=\gamma^0{\gamma^{\hat\mu}}^+\gamma^0$, and satisfy the anti-commutation rules $\{ \gamma^{\hat\alpha},\, \gamma^{\hat\beta} \}=2\eta^{\hat\alpha \hat\beta}$ giving the $SL(2,{\Bbb C})$ generators as, 
\begin{equation}\label{Sab}
S^{\hat\alpha \hat\beta}=\frac{i}{4}\, [\gamma^{\hat\alpha},
\gamma^{\hat\beta} ]\,.
\end{equation}
The notation $\,^+$ stands for the Hermitian conjugation of matrices that has to do not be  confused with the Hermitian conjugations with respect to the relativistic scalar products we will introduced later. 

Mathematically speaking,  the Dirac spinors transform according to the reducible rep. $\rho_D= (\frac{1}{2},0)\oplus (0,\frac{1}{2})$ of the $SL(2,{\Bbb C})$ group but which is the simpler rep. allowing invariant bilinear forms. For this reason we consider here that $\rho_D$ is just the rep. defining this group denoting its elements simply as $\rho_D(A)\equiv A\in SL(2,{\Bbb C})$ \cite{CD6}. We remind the reader that in the covariant parametrization with skew-symmetric parameters, $\hat\omega_{\hat\alpha\hat\beta}=-\hat\omega_{\hat\beta\hat\alpha}$, the $SL(2,{\Bbb C})$ transformations 
\begin{equation}
A(\hat\omega)=\exp\left(-\frac{i}{2}\,\hat\omega_{\hat\alpha\hat\beta}S^{\hat\alpha\hat\beta}\right)\,,
\end{equation}
correspond through the canonical homomorphism to the transformation matrices
$\Lambda[A(\hat\omega)]=\Lambda(\hat\omega)\in L^{\uparrow}_{+}$ having the matrix elements $\Lambda(\hat\omega)^{\hat\alpha\,\cdot}_{\cdot \,\hat\beta}=\delta^{\hat\alpha}_{\hat\beta}+\hat\omega^{\hat\alpha\,\cdot}_{\cdot \,\hat\beta}+...$.

The covariance under the gauge transformation (\ref{gauge1}) and (\ref{gauge2}) produced by $\Lambda[A(x)]\in L^{\uparrow}_{+}$, when the Dirac field transforms as
\begin{equation}
\psi(x)\to  \psi'(x)=  A(x)\psi(x)\,,\label{Gauge}
\end{equation}
is assured by  the covariant derivatives 
\begin{equation}
\nabla_{\hat\alpha}=e_{\hat\alpha}+\frac{i}{2}\hat\Gamma^{\hat\gamma}_{\hat\alpha
\hat\beta}S^{\hat\beta\,
\cdot}_{\cdot\, \hat\gamma}\,,
\end{equation}
where 
$\hat\Gamma^{\hat\sigma}_{\hat\mu \hat\nu}=e_{\hat\mu}^{\alpha} e_{\hat\nu}^{\beta}
(\hat e_{\gamma}^{\hat\sigma}\Gamma^{\gamma}_{\alpha \beta} -\hat e^{\hat\sigma}_{\beta, \alpha})$ are the connection components in local frames (known as the spin connections) expressed in terms of tetrads and Christoffel symbols,   $\Gamma^{\gamma}_{\alpha \beta}$. From the action (\ref{action}) it results the Dirac equation 
\begin{equation}\label{EEE}
\left(  i\gamma^{\hat\alpha}\nabla_{\hat\alpha}-m\right)\psi(x)=0\,,
\end{equation}
that can be written in explicit form as
\begin{equation}\label{Dir2}
i\gamma^{\hat\alpha}e_{\hat\alpha}^{\mu}\partial_{\mu}\psi - m\psi
+ \frac{i}{2} \frac{1}{\sqrt{g}}\partial_{\mu}(\sqrt{g}e_{\hat\alpha}^{\mu})
\gamma^{\hat\alpha}\psi -\frac{1}{4}
\{\gamma^{\hat\alpha}, S^{\hat\beta \cdot}_{\cdot \hat\gamma} \}
\hat\Gamma^{\hat\gamma}_{\hat\alpha \hat\beta}\psi =0\,.
\end{equation}
The particular solutions of this equation form a vector space equipped with the conserved relativistic scalar product  \cite{CD1}
\begin{equation}\label{SPD}
\langle \psi, \psi'\rangle=\int_{\Sigma} d\sigma_{\mu} \sqrt{g}\, e^{\mu}_{\hat\alpha}
\,\bar{\psi}(x)\gamma^{\hat\alpha}\psi'(x)\,, 
\end{equation}
whose integral is performed on a space-like section $\Sigma\subset M$. 

Applying these general formulas to the Dirac field in the frames $\{t,{\bf x};e\}$ of a FLRW manifold $(M,a)$ we obtain the Dirac equation
\begin{equation}\label{ED1}
\left(i\gamma^0\partial_{t}+i\frac{1}{\alpha(t)}\gamma^i\partial_i
+\frac{3i}{2}\frac{\dot{\alpha}(t)}{\alpha(t)}\gamma^{0}-m\right)\psi(t,{\bf x})=0\,,
\end{equation}
while in the frames $\{t,{\bf \underline{x}};e\}$ this takes the form
\begin{equation}\label{ED2}
\left[i\gamma^0\partial_{t}+i{\gamma^i}{\underline{\partial}_{\,i}} -m
+i\gamma^{0}\frac{\dot{\alpha}(t)}{\alpha(t)}
\left(\underline{x}^i\underline{\partial}_{\,i}+\frac{3}{2}\right)\right]\psi(t,{\bf \underline{x}})=0\,.
\end{equation}
Similar results can be written in the frames of conformal time performing the substitution (\ref{subs}) in the above equations. The different versions of the Dirac equation in  frames with spherical coordinates will be analysed  later when the spherical modes will be studied.  

\subsection{Conserved observables}

The general relativistic covariance under diffeomorphisms and gauge transformations is not able to generate conserved quantities via Noether theorem. Only the isometries can do that and for this reason these deserves a special attention. 

In general, the space-times  $(M,g)$ of physical interest have isometries, $x\to x'=\phi_{\frak g} (x)$, which are  non-linear transformations  preserving the metric. These form the isometry group $I(M)$ having the composition rule $\phi_{\frak g}\circ \phi_{{\frak g}'}=\phi_{\frak{gg}'}$, $\forall \frak{g,g}'\,\in I(M)$ and  the identity function $id=\phi_{\frak e}$ as the unit element. In a given parametrization, ${\frak g}={\frak g}(\xi)$ (with ${\frak e}={\frak g}(0)$), the isometries
\begin{equation}
x\to x'=\phi_{{\frak g}(\xi)}(x)=x+\xi^a k_a(x) +... 
\end{equation}
lay out the Killing vectors $k_a=\partial_{\xi_a}\phi_{{\frak g}(\xi)}|_{\xi=0}$ associated to the parameters $\xi^a$ ($a,b,...=1,2...N$).  

Since the isometries may change the relative position of the natural and local frames  we proposed the theory of external symmetry \cite{ES,dS} introducing the combined transformations $(A_{\frak g},\phi_{\frak g})$  able to correct the positions of the local frames after each isometry. These transformations must preserve not only the metric  but the tetrad gauge too, transforming  the 1-forms as $\tilde\omega(x')=\Lambda[A_{\frak g}(x)]\tilde\omega(x)$. Hereby, it results the form of the local transformations   \cite{ES},
\begin{equation}\label{Axx}
\Lambda^{\hat\alpha\,\cdot}_{\cdot\,\hat\beta}[A_{\frak g}(x)]= \hat
e_{\mu}^{\hat\alpha}[\phi_{\frak g}(x)]\frac{\partial
\phi^{\mu}_{\frak g}(x)} {\partial x^{\nu}}\,e^{\nu}_{\hat\beta}(x)\,,
\end{equation}
which define the matrices $A_{\frak g}(x)$ assuming, in addition, that  $A_{{\frak g}={\frak e}}(x)=1\in SL(2,\Bbb C)$. The resulted combined transformations $(A_{\frak g},\phi_{\frak g})$ preserve the gauge,  $e'=e$ and $\tilde\omega'=\tilde\omega$,  transforming the  Dirac field according to the covariant rep.  $T : (A_{\frak g},\phi_{\frak g})\to T_{\frak g}$   whose operators act as
\begin{equation}\label{Tx}
(T_{\frak g}\psi)[\phi_{\frak g}(x)]=A_{\frak g}(x)\psi(x)\,.
\end{equation}
We have shown that the pairs $(A_{\frak g},\phi_{\frak g})$ constitute a well-defined Lie group with respect to the new operation  that can be seen as a rep. of the universal covering group of $I(M)$ denoted here by $S(M)$ \cite{ES}. In fact, these covariant reps. transfer the Lorentz covariance from special relativity to general relativity. 

Given a parametrization, ${\frak g}={\frak g}(\xi)$, for small values of  $\xi^{a}$, we may expand the parameters of the transformation $A_{{\frak g}(\xi)}(x)\equiv A[\hat\omega_{\xi}(x)]$  as
$\hat\omega^{\hat\alpha\hat\beta}_{\xi}(x)= \xi^{a}\Omega^{\hat\alpha\hat\beta}_{a}(x) +\cdots$, in terms of the functions
\begin{equation}\label{Om}
\Omega^{\hat\alpha\hat\beta}_{a}\equiv {\frac{\partial
\hat\omega^{\hat\alpha\hat\beta}_{\xi}} {\partial\xi^a}}_{|\xi=0}
=\left( \hat e^{\hat\alpha}_{\mu}\,k_{a,\nu}^{\mu} +\hat
e^{\hat\alpha}_{\nu,\mu}
k_{a}^{\mu}\right)e^{\nu}_{\hat\lambda}\eta^{\hat\lambda\hat\beta}
\end{equation}
that  are skew-symmetric,
$\Omega^{\hat\alpha\hat\beta}_{a}=-\Omega^{\hat\beta\hat\alpha}_{a}$,
only if $k_a$ are Killing vectors. In this case we obtain the  basis-generators
of the covariant reps. \cite{ES},
\begin{equation}\label{Xa}
X_{a}=\left.i\partial_{\xi^a}T_{{\frak g}(\xi)}\right|_{\xi=0}=-i
k_a^{\mu}\partial_{\mu} +\frac{1}{2}\,\Omega^{\hat\alpha\hat\beta}_{a}
S_{\hat\alpha\hat\beta}\,,
\end{equation}
which  can be put in covariant form \cite{CML,CCML}. These  generators  are the principal conserved observables of the quantum theory which commute with the operator of the Dirac equation.  

In a given frame, any conserved operator $X$  gives rise to the conserved quantity 
$C[X]=\langle\psi,X\psi\rangle$ derived with the help of the scalar product (\ref{SPD}). This quantity is interpreted as the expectation value at the level of the relativistic quantum mechanics and after the second quantization  becomes the corresponding one particle operators of the QFT \cite{CD6,CD7}. We have shown that the relativistic scalar product (\ref{SPD}) is  invariant under isometries, $\langle T_{\frak g}\psi,T_{\frak g}\psi'\rangle=\langle\psi,\psi'\rangle$, while all  the conserved observables  (\ref{Xa})  are self-adjoint with respect to this scalar product,  $\langle X_a\psi,\psi'\rangle=\langle\psi,X_a\psi'\rangle$. 

The spatially flat FLRW space-times, $(M,a)$,  have, in general, the Euclidean  isometry group $E(3)$ formed by space translations and space rotations. The basis-generators of these isometries  are  the components $P^i$ of the momentum operator and those of the total angular momentum $J_i$. In the charts $\{t, {\bf x}\}$ and $ \{t, \underline{\bf x}\}$ the momentum components can be written as
\begin{equation}
P^i =-i \,\partial_i=-i\, a(t)\, \underline{\partial}_{\,i}\,,
\end{equation}
according to Eq. (\ref{X1}), while those of the total  angular momentum read, 
\begin{equation}
J_i=L_i+S_i\,, \quad S_i =\frac{1}{2}\,\varepsilon_{ijk}S_{jk}\,,
\end{equation}
where $L_i=-i \epsilon_{ijk}x^i \partial_k=-i \epsilon_{ijk}\underline{x}^i \underline{\partial}_{\,k}$ are the components of the orbital angular momentum which is not conserved in the Dirac theory. The conserved observables, $P^i$ and $J_i$, can generate freely many other conserved observables as, for example, the Pauli-Lubanski operator 
\begin{equation}\label{PaLu}
W={\bf P}\cdot{\bf J}={\bf P}\cdot{\bf S}\,,
\end{equation}
which allows one to define the helicity (as in Appendix A). 

A special case is the dS space-time  having the scale factor 
\begin{equation}\label{Hubb}
a(t)=e^{\omega t}\to ~~\frac{\dot{a}(t)}{a(t)}=\omega\,,
\end{equation}
where $\omega$ is the Hubble dS constant in out notation. The  dS isometry group $SO(1,4)$ lays out ten basis-generators,  $P^i$,  $J_i$,  three more Abelian generators $Q^i$ and the energy operator which has different forms,
\begin{equation}\label{HdS}
H = i\partial_t + \omega\, {\bf X}\cdot {\bf P}\,,\quad \underline{H}=i \partial_t\,,
\end{equation}
in the local charts $\{t,{\bf x}\}$ and respectively $\{t,{\bf \underline{x}}\}$ \cite{CGRG,CD3}.
${\bf X}$ is the usual multiplicative operator acting as $(X^i f)({\bf x})=x^i f({\bf x})$. 

\subsection{Representations}

The Dirac theory lays out the natural $U_{em}(1)$ internal symmetry, giving rise to the conserved electromagnetic current density $\overline{\psi}\gamma^{\hat\mu}\psi$ whose sign may be changed by the  charge conjugation  $\psi\to \psi^c={\cal C}\psi^*$, with ${\cal C}=i\gamma^2$,  since the $\gamma$-matrices satisfy 
${\cal C}{\gamma^{\hat\mu}}^*{\cal C}=-\gamma^{\hat\mu}$.  Consequently, for every particular solution $U$ of the Dirac equation there exists the charge conjugated solution $V={\cal C}U^*$ regardless the geometry of the background.  Thus the space of the solutions is split into the set of solutions of positive frequencies associated to particles and the set of their charge conjugated solutions which are of negative frequencies describing antiparticles. The problem  is how a such basis can be defined in the space of solutions, separating the frequencies without ambiguities.

The traditional method is to look for a complete system of commuting conserved observables $\{A\}=\{ A_1,A_2,...A_n\}$  able to determine a (generalized) basis formed by the solutions of the Dirac equation which, in addition, solve the common eigenvalue problems
\begin{equation}\label{separ}
A_i U_{\alpha}=a_i U_{\alpha}\,, \quad A_i V_{\alpha}=\pm a_i V_{\alpha}\,,\quad i=1,2...n\,.
\end{equation}
corresponding to the eigenvalues $\alpha=\{a_1,a_2,...a_n\}$ that form  the spectrum ${\Bbb S}={\Bbb S}_d\cup{\Bbb S}_c$ which may have a discrete part ${\Bbb S}_d$ and a continuous one ${\Bbb S}_c$. In this manner all the integration constants get physical meaning as eigenvalues of these operators that have a precise physical interpretation.    

When the system of observables is complete, this determines a basis of the space of solutions said to be of the rep. $\{A\}$. Unfortunately, in the case of our geometries we do not find such complete systems of operators since the isometry groups of the FLRW space-times, including the dS one, do not have Cartan sub-algebras with more than three generators while for completing such systems we need at least four generators, as the components of the four-momentum operator of the Minkowski space-time. Thus we must make do with incomplete systems of three operators resorting to supplemental hypotheses for setting all the integration constants we need for separating the frequencies, determining thus the vacuum.  

In a given  rep. $\{A\}$ the Dirac field can be expanded  as  
\begin{equation}\label{psigen}
\psi(x)=\psi^{(+)}(x)+\psi^{(-)}(x)=\int_{\alpha\in{\Bbb S}}U_{\alpha}(x)a(\alpha)+V_{\alpha}(x) b^{*}(\alpha)\,,
\end{equation}
where we sum over the discrete part ${\Bbb S}_d$ and integrate over the continuous part ${\Bbb S}_c$ of the spectrum ${\Bbb S}$. The functions $a$ and $b$ are the particle and respectively antiparticle wave functions of the rep. $\{A\}$. Under canonical quantization, these functions become field operators, $a\to {\frak a}$ and $b^*\to {\frak b}^{\dagger}$,   satisfying  the canonical anti-commutation relations \cite{BDR,CD7} 
\begin{eqnarray}
&&\left\{a(\alpha),a^{\dagger}(\alpha')\right\}=\left\{b(\alpha),b^{\dagger}(\alpha')\right\}\nonumber\\
&&\hspace*{27mm}=\delta(\alpha,\alpha')=\left\{
\begin{array}{ll}
\delta_{\alpha,\alpha'}\,,& \alpha,\alpha'\in {\Bbb S}_d\\
\delta(\alpha-\alpha')\,,& \alpha,\alpha'\in {\Bbb S}_c
\end{array}\right.\,,\label{cac}
\end{eqnarray}
requested by the Fermi-Dirac statistics, while  the fundamental spinors $U_{\alpha}$ and $V_{\alpha}$ have to form an orthonormal basis of the rep. $\{A\}$ complying with corresponding orthonormalization relations. For example, in the chart $\{t, {\bf x}\}$ these spinors are orthonormal,
\begin{eqnarray}
\left<U_{\alpha},U_{\alpha'}\right>&=&
\left<V_{\alpha},V_{\alpha'}\right>=
\delta(\alpha,\alpha')
\label{orto1a}\\
\left<U_{\alpha},V_{\alpha'}\right>&=&
\left<V_{\alpha},U_{\alpha'}\right>=
0\,,\label{orto2a}
\end{eqnarray} 
with respect to the scalar product (\ref{SPD1}) and satisfy the completeness condition
\begin{equation}\label{complet}
\int_{\alpha\in{\Bbb S}}U_{\alpha}(t,{\bf x})\,\overline{U}_{\alpha}(t,{{\bf x}\,}')+V_{\alpha}(t,{\bf x})\,\overline{V}_{\alpha}(t,{{\bf x}\,}')=\frac{1}{a(t)^3}\,\delta^3({\bf x}-{{\bf x}\,}')\,,
\end{equation}
associated to this scalar product \cite{CD7}. 

In what follows we focus only on various solutions of the Dirac equation remaining at the level of the relativistic quantum mechanics where the Dirac field $\psi$ depends on  the wave functions $a$ and $b$ that can be pointed out by using the inversion formulas
\begin{equation}\label{invers}
a(\alpha)=\left<U_{\alpha}, \psi\right>\,,\quad
b(\alpha)=\left<\psi, V_{\alpha}\right>\,.
\end{equation} 
Note that these  functions transform under isometries alike according to a unitary rep. of the isometry group such that  Eq. (\ref{psigen}) can be seen as defining the equivalence between the covariant rep.  (\ref{Tx}) and the unitary one transforming the functions $a$ and $b$ \cite{CGRG,CD6,CD7}. 

\section{Time evolution pictures}

Working with different local charts  we may face with some difficulties since  at the quantum level the time-dependent coordinate transformations (\ref{xx}) are not compatible with the usual quantum formalism where it is easier to change the time evolution pictures rather than transforming the coordinates as ${\bf x}\to {\bf \underline{x}}$. For this reason we defined two different time evolution pictures which prevent us to work with such coordinate transformations, remaining with an unique set of space coordinates, ${\bf x}$. 

\subsection{Related pictures}

The first picture is the mentioned NP which  is just the usual  theory in the frame $\{t,{\bf x};e\}$ where the Dirac field  $\psi$ satisfies Eq.  (\ref{ED1}) that can be rewritten as
\begin{equation}\label{EPPN}
\left({E}_D(t)-m\right)\psi(t,{\bf x})=0
\end{equation}
pointing out its operator 
\begin{equation}\label{EDNP}
{E}_D(t)=i\gamma^0\partial_{t}
+\frac{3i}{2}\frac{\dot{\alpha}(t)}{\alpha(t)}\gamma^{0}-\frac{1}{\alpha(t)}{\bf \gamma}\,\cdot{\bf P}\,.
\end{equation}
which depends explicitely on time apart from the dS case when the Hubble function (\ref{Hubb}) becomes constant.  The scalar product derived from Eq. (\ref{SPD}) reads   \begin{equation}\label{SPD1}
\langle \psi, \psi'\rangle=\int d^3 x\, a(t)^3 \,\bar{\psi}(x)\gamma^{0}\psi'(x)\,. 
\end{equation}

The second picture is the SP which governs the time evolution of the new field \cite{CD3},
\begin{equation}\label{defS}
 \psi_S(t,{\bf x})=\psi \left(t,\frac{1}{a(t)}{\bf x}\right)\,,
\end{equation}
according to Eq.  (\ref{ED2}) in which we substitute 
\begin{equation}\label{subx}
{\bf \underline{x}}\to {\bf x}\,, \quad \underline{\partial}_{\,i}\to \partial_i\,,
\end{equation}
obtaining the Dirac equation of SP,  
\begin{equation}\label{EPPS}
\left({E}_D^S(t)-m\right)\psi_S(t,{\bf x})=0\,.
\end{equation}
with the new operator 
\begin{equation}\label{EDSP}
{E}_D^{S}(t)=i\gamma^0\partial_{t} 
-\gamma^{0}\frac{\dot{\alpha}(t)}{\alpha(t)}
\left({\bf X}\cdot{\bf P}-\frac{3i}{2}\right)-{\bf \gamma}\cdot{\bf P}\,.
\end{equation}
The relativistic scalar product of SP can be derived after performing the substitution (\ref{subx}) in Eq. (\ref{SPD}) obtaining the expression \cite{CD3}
\begin{equation}\label{SPSP}
\langle \psi, \psi'\rangle_S=\int d^3 x\,  \,\bar{\psi}(t,{\bf x})\gamma^{0}\psi'(t,{\bf x})\,, 
\end{equation}
which has the same form as in $(M,\eta)$. Therefore, the scalar products of these pictures are different generating two different types of Hermitian conjugations denoted by $^{\dagger}$ for NP and by $^{\ddagger}$ for SP.

The advantage of this approach is of working with only one set  of coordinate and momentum operators, ${\bf X}$ and ${\bf P}$, which  satisfy the canonical commutation rule $[X^i  ,P  ^j]=i\delta_{ij}$. These operators are Hermitian with respect to both the relativistic scalar products  of NP and SP,  ${\bf X}^{\dagger}={\bf X}^{\ddagger}={\bf X}$, and  ${\bf P}^{\dagger}={\bf P}^{\ddagger}={\bf P}$. With their help, and with the Dirac matrices $\gamma^{\hat\alpha}$, we can generate freely the operator algebra  ${\cal A}({\bf X},{\bf P},\gamma)$ of our pictures, constituted by all the analytic functions of the mentioned operators and matrices \cite{CD3,CD7}.  

Of a special interest is the dilation generator,
\begin{equation}\label{Dil}
D= {\bf X}  \cdot{\bf P}\,, 
\end{equation}
which satisfies the commutation rules $\left[D,X^i\right]=-iX^i$ and $\left[D,P^i\right]=i P^i$. This operator  is  non-Hermitian, $D^{\dagger}=D^{\ddagger}=D-3i$, but the operator 
\begin{equation}\label{DDD}
\hat D=D-\frac{3i}{2}\,, 
\end{equation}
which satisfies the same commutation rules has this property, $\hat D={\hat D}^{\dagger}={\hat D}^{\ddagger}$. Consequently, the Dirac operator of the SP is Hermitian, $E_D^S(t)^{\ddagger}= E_D^S(t)$ while that of NP does not have this property since 
\begin{equation}\label{EDdagg}
 {E}_D(t)^{\dagger}= {E}_D(t)+3i\frac{\dot{\alpha}(t)}{\alpha(t)}\gamma^{0}\,,
\end{equation}
as it results from Eq. (\ref{EDNP}).  

For relating the above defined pictures we need a transformation operator  $ T: NP\to SP$ acting as $A_S=TAT^{-1}$ for any operators $A\in NP$ and $A_S \in SP$.  We observe that the transformation \cite{CD3}
\begin{equation}\label{U}
T(t)=e^{-i\ln(a(t)) D}\,,
\end{equation}
generated by the dilation operator (\ref{Dil}), can take over the role of the coordinates transformation (\ref{xx}) transforming any analytic function $F({\bf X})$ or $G({\bf P})$  as
\begin{eqnarray}
T(t)F( {\bf X})T(t)^{-1}&=&F\left(\frac{1}{a(t)}{\bf X}\right)\,,\\
T(t)G({\bf P})T(t)^{-1}&=&G\left(a(t){\bf P}\right)\,.
\end{eqnarray}
Therefore, we may write the desired transformation  
\begin{equation}
\psi_S(t,{\bf x})=T(t)\psi(t,{\bf x})=\psi \left(t,\frac{1}{a(t)}{\bf x}\right)\,,
\end{equation}
which is in accordance with Eq. (\ref{defS}) and defines the coordinate and momentum operators of SP as, 
\begin{eqnarray}
{\bf X}_S&=&T(t){\bf X}T(t)^{-1}=\frac{1}{a(t)}{\bf X} \in SP\,, \\
{\bf P}_S&=&T(t){\bf P}T(t)^{-1}={a(t)}{\bf P} \in SP\,.
\end{eqnarray}
In addition, we obtain the important transformation rule 
\begin{equation}
T(t) i\partial_t T(t)^{-1}=i\partial_t-\frac{\dot{a}(t)}{a(t)} D\,,
\end{equation}
which enables us to relate  the operators of the Dirac equations of our pictures as
\begin{equation}
{E}_D^S(t)=T(t) {E}_D(t) T(t)^{-1}\,,
\end{equation}
pointing out their equivalence. 

Obviously,  this equivalence is not a unitary one  since the transformation operator (\ref{U}) is non-unitary. Indeed, by using Eq. (\ref{EDdagg}) it is not difficult to deduce the following rule \cite{CD3}
\begin{equation}
T(t)^{\dagger}=T(t)^{\ddagger}=a(t)^3 T(t)^{-1}\,,
\end{equation}
which shows that $T(t)$ is non-unitary. This is not an impediment since we can obtain an  unitary equivalence  if we replace NP by its Minkowskian projection we will discuss later.
 
\subsection{Energy and Hamiltonian operators}

The energy operator is conserved only in two particular spatially flat FLRW space-times, namely the Minkowski and the dS manifolds which have larger isometry groups, the Poincar\' e group and respectively the $SO(1,4)$ ones. The energy operator of the Minkowski geometry is just the time-like component of the four momentum, $P^0=i\partial_t$. 

In the case of the dS geometry the energy operator is a Killing vector field which is time-like only inside the null cone of an observer staying at rest in origin \cite{CGRG}. This takes different forms depending on the local coordinates and implicitly on the time evolution picture we adopt. In  SP the dS energy operator has the simpler form 
\begin{equation}\label{hsp}
\underline{H}\to H_S=i\partial_t\,,
\end{equation}
as it results from Eq. (\ref{HdS}). 

In what follows  we generalize this result  assuming that the energy operator is defined by  Eqs. (\ref{hsp}) in SP of any FLRW space-time $(M,a)$. Consequently,  in NP  the energy operator becomes
\begin{equation}\label{hnp}
H=T(t)^{-1}H_ST(t)=i\partial_t +\frac{\dot{a}(t)}{a(t)}D\,,
\end{equation}
having the algebraic properties
\begin{equation}\label{HXP}
\left[H, {\bf X}\right]=-i\frac{\dot{a}(t)}{a(t)}{\bf X}\,,\quad \left[H, {\bf P}\right]=i\frac{\dot{a}(t)}{a(t)}{\bf P}\,.
\end{equation}
Then, bearing in mind that $D$ is related to the Hermitian operator $\hat D$ as in  Eq. (\ref{DDD}) and observing that 
\begin{equation}
(i\partial_t)^{\ddagger}=i\partial_t\,, \quad  (i\partial_t)^{\dagger}=i\partial_t +3\frac{\dot{a}(t)}{a(t)}
\end{equation}
we understand that both these operators are Hermitian, $H=H^{\dagger}$ and $H_S=H^{\ddagger}_S$, with respect to the specific scalar products of their pictures, (\ref{SPD1}) and  (\ref{SPSP}). 
 
The above defined energy operators give the related eigenvalues problems
\begin{equation}
H F=E F\,, \quad H_S F_S=E F_S\,,
\end{equation}
which are solved by the eigenfunctions, 
\begin{eqnarray}
F_S(t, {\bf x})&=&f({\bf x})e^{-iEt}\\ 
F(t,{\bf x})&=&T(t)^{-1}F_S(t,{\bf x})=f\left(a(t){\bf x}\right) e^{-iEt}\,,
\end{eqnarray}
where $E\in {\Bbb R}$ is the energy while $f$ is an arbitrary function of ${\bf x}$. In general, these eigenfunctions cannot be solutions of the Dirac equation since the energy operators do not commute with ${E}_D$ or ${E}_D^S$  apart from the dS case when the energy is conserved. 

Now we may define the Hamiltonian operators of our pictures allowing us to bring the Dirac equations (\ref{EPPN}) and (\ref{EPPS}) in the Schr\" odinger form, 
\begin{eqnarray}
&NP:&\quad H \psi (t,{\bf x})={H}_D(t)\psi(t,{\bf x})\,,\label{HDNP}\\
&SP:&\quad H_S\psi_S(t,{\bf x})={H}_D^S(t)\psi_S(t,{\bf x})\,. \label{HDSP}
\end{eqnarray}
According to Eqs. (\ref{EDNP}) and (\ref{EDSP}) we obtain the related Hamiltonian operators, ${H}_D^S(t)=T(t){H}_DT(t)^{-1}$, that read 
\begin{eqnarray}
{H}_D(t)&=&\frac{1}{a(t)}\gamma^0\gamma^iP^i+\gamma^0 m +\frac{\dot{a}(t)}{a(t)}\hat D\nonumber\\
&=&{H}_D'(t)+{H}_{\rm int}(t)\\
{H}_D^S(t)&=&\gamma^0\gamma^iP^i+\gamma^0 m +\frac{\dot{a}(t)}{a(t)}\hat D\nonumber\\
&=&{H}_{D}^0+{H}_{\rm int}(t)
\end{eqnarray}
where $\hat D$ is the Hermitian dilation generator  (\ref{DDD}).  

Similar time evolution pictures can be defined in the charts with conformal time by substituting $t\to t_c$ according to Eqs. (\ref{subs}) in the expressions of all the operators considered here. Thus we obtain the new pictures namely, the natural picture with conformal time, $NP_c$, and its associated Schr\" odinger picture,  $SP_c$. 

It is remarkable that in SP we obtain a  typical structure of a problem of perturbations since  the Hamiltonian is formed by the usual  Hamiltonian of the free Dirac field on Minkowski space-time, ${H}_D^0=\gamma^0\gamma^iP^i+\gamma^0 m$,  and the time-dependent interaction Hamiltonians due to the background gravity,
\begin{equation}\label{Hint}
{H}_{\rm int}=\frac{\dot{a}(t)}{a(t)}\,\hat D \in SP\,,\quad {H}_{{\rm int}\,c}=\frac{\dot{a}(t_c)}{a(t_c)^2}\,\hat D \in SP_c\,,
\end{equation}
which are proportional to the Hubble function. Thus the unperturbed problem is the Dirac free field on the Minkowski space-time while the gravitational affect is encapsulated in the interaction Hamiltonian. 

In NP we do not have this opportunity since the energy operator has the unusual form (\ref{hnp}), depending on the Hubble function, while ${H}_D'(t)$  cannot be interpreted as an unperturbed Hamiltonian as long as this depends explicitely on $a(t)$. This picture, in which the momentum operator is simpler, is suitable for studying the plane or spherical wave solutions depending on the vector or scalar momentum.   

\subsection{Minkowskian projection}

Technically speaking there is a transformation of the fields and operators of NP 
\begin{equation}\label{NPMP}
\psi\to \hat\psi=a(t)^{\frac{3}{2}}\psi\,, \quad X\to \hat X =a(t)^{\frac{3}{2}}X\,a(t)^{-\frac{3}{2}}\,,
\end{equation}
which brings the  Dirac operator in the simpler form
\begin{equation}\label{EqDpr}
\hat E_D(t)=i\gamma^0\partial_{t}-\frac{1}{a(t)}\,{\bf \gamma}\cdot {\bf P}\,,
\end{equation}
while the scalar product becomes identical to that of SP  since, 
\begin{equation}\label{SPMP1}
\langle \psi,\psi'\rangle =\langle \hat\psi, \hat\psi'\rangle_S\,, 
\end{equation}
taking the same form as in $(M,\eta)$. For this reason we say that this is the Minkowskian projection (MP) of NP observing that this offers us the advantages of a simpler Dirac equation and a common scalar product. Now  the Hermitian conjugation of the MP is similar to that of SP and, consequently, will be denoted alike with $\,^{\ddagger}$. The Dirac operator (\ref{EqDpr}) is now Hermitian with respect to this scalar product. The transformations (\ref{NPMP}) do not affect the coordinate and momentum operators but change the energy operator, 
\begin{equation}
H\to \hat H =H-\frac{3i}{2}\frac{\dot{a}(t)}{a(t)}=i\partial_t +\frac{\dot{a}(t)}{a(t)}\hat D\,,
\end{equation}
which depends now on the Hermitian dilation operator (\ref{DDD}) instead of $D$ as in Eq. (\ref{hnp}). Therefore, the operator $\hat H$ is Hermitian with respect to the scalar product (\ref{SPSP}).

The principal feature of the MP is that this is equivalent to SP through  the unitary  transformation $\psi_{S}(t,{\bf x})=U(t)\hat \psi(t,{\bf x})$ whose operator
\begin{equation}\label{UD}
U(t)=a(t)^{-\frac{3}{2}}T(t)=e^{-i\ln(a(t)) \hat D}\,, 
\end{equation}
generated now by the Hermitian dilation operator,  is unitary with respect to the  scalar product (\ref{SPD}). 

Another opportunity is of comparing the states of NP of $(M,a)$ with the states prepared in the Minkowski space-time $(M,\eta)$.  This can be done  if we chose the {\em same} coordinates on both these manifolds since the scalar products giving the quantities with physical meaning are similar. Note that this choice  is possible at any time since the manifolds $(M,a)$ are local Minkowkian. Therefore, given a Dirac states whose spinor in the MP is $\hat \psi$ and a spinor $\psi_M$ of a Minkowski state we can construct the time dependent quantity  
\begin{equation}
{\cal P}(t)=\left |\left<\hat\psi(t),\psi_M(t)\right>_S\right|^2\,,
\end{equation}
by using the scalar product (\ref{SPD}). This quantity can be interpreted as the probability of measuring at the time $t$ the parameters of the state $\psi_M$ in the state $\hat\psi$ prepared in $(M,a)$. With their help we can imagine detectors measuring parameters of Minkowskian states on FLRW space-times which may be helpful auxiliary tools in  refining the physical interpretation of the quantum modes.   

\section{Representations in NP}

As mentioned, the conserved generators ${\bf P}$ and ${\bf J}$ of the $E(3)$ isometries of the FLRW manifolds $(M,a)$ are not able to generate  freely an  algebra rich enough for selecting complete systems of commuting operators.  Therefore, we must make do with the incomplete system  giving the plane waves of the momentum-helicity or those of the momentum-spin reps. and with the spherical waves of the total angular momentum rep.. 

\subsection{Plane P-waves}

In the frames $\{t,{\bf x};e\}$ of NP of the space-times $(M,a)$ the general solution of the Dirac equation (\ref{EPPN}) may be written as a mode integral, 
\begin{eqnarray}
\psi(t,{\bf x}\,)& =& 
\psi^{(+)}(t,{\bf x}\,)+\psi^{(-)}(t,{\bf x}\,)\nonumber\\
& =& \int d^{3}p
\sum_{\sigma}[U_{{\bf p},\sigma}(x)a({\bf p},\sigma)+V_{{\bf p},\sigma}(x){b}^{*}({\bf p},\sigma)]\,,\label{p3}
\end{eqnarray}
in terms of the fundamental spinors $U_{{\bf p},\sigma}$  and  $V_{{\bf p},\sigma}$ of positive and respectively negative frequencies which are plane waves solutions of the Dirac equation (\ref{EPPN}), defined as common eigenspinors of the momentum components  $\{P^1,P^2,P^3\}$ corresponding to the eigenvalues $\{p^1,p^2,p^3\}$ representing  the components of the  conserved momentum ${\bf p}$. In addition, these solutions  depend on a polarization $\sigma$  that can be defined in different manners as we show in the Appendix A. 

These spinors  form an orthonormal basis with respect to the scalar product (\ref{SPD1})  being related  through the charge conjugation, 
\begin{equation}\label{chc}
V_{{\bf p},\sigma}(t,{\bf x})=U^c_{{\bf p},\sigma}(t,{\bf x}) =i\gamma^2\left[{U}_{{\bf p},\sigma}(t,{\bf x})\right]^* \,,
\end{equation}
and satisfying the orthogonality relations
\begin{eqnarray}
\langle U_{{\bf p},\sigma}, U_{{{\bf p}\,}',\sigma'}\rangle &=&
\langle V_{{\bf p},\sigma}, V_{{{\bf p}\,}',\sigma'}\rangle=
\delta_{\sigma\sigma^{\prime}}\delta^{3}({\bf p}-{\bf p}\,^{\prime})\label{ortU}\\
\langle U_{{\bf p},\sigma}, V_{{{\bf p}\,}',\sigma'}\rangle &=&
\langle V_{{\bf p},\sigma}, U_{{{\bf p}\,}',\sigma'}\rangle =0\,, \label{ortV}
\end{eqnarray}
and a completness condition of the form (\ref{complet}). This basis defines the momentum rep. that depends, in addition, on the way in which the polarization is defined.

In the standard rep. of the Dirac matrices (with diagonal $\gamma^0$) the general form of the fundamental spinors in momentum rep.,   
\begin{eqnarray}
U_{{\bf p},\sigma}(t,{\bf x}\,)&=&\frac{e^{i{\bf p}\cdot{\bf x}}}{[2\pi a(t)]^{\frac{3}{2}}}\left(
\begin{array}{c}
u^+_p(t) \,
\xi_{\sigma}\\
u^-_p(t) \,
{\bf \sigma}\cdot{\bf n}_p\,\xi_{\sigma}
\end{array}\right)\,,
\label{Ups}\\
V_{{\bf p},\sigma}(t,{\bf x}\,)&=&\frac{e^{-i{\bf p}\cdot{\bf x}}}{[2\pi a(t)]^{\frac{3}{2}}} \left(
\begin{array}{c}
v^+_p(t)\,
{\bf \sigma}\cdot{\bf n}_p\,\eta_{\sigma}\\
v^-_p(t) \,\eta_{\sigma}
\end{array}\right)
\,,\label{Vps}
\end{eqnarray}
is determined by the time modulation functions (t.m.f.) $u^{\pm}_p(t)$ and $v^{\pm}_p(t)$  that depend only on $t$ and  $p=|{\bf p}|$.  The notation ${\bf n}_p$ stands for the unit vector of the momentum direction while  $\xi_{\sigma}$ and $\eta_{\sigma}= i\sigma_2 (\xi_{\sigma})^{*}$ are Pauli spinors supposed to be correctly normalized,  $\xi^+_{\sigma}\xi_{\sigma'}=\eta^+_{\sigma}\eta_{\sigma'}=\delta_{\sigma\sigma'}$. In addition, they must satisfy the completeness condition 
\begin{equation}\label{Pcom}
\sum_{\sigma}\xi_{\sigma}\xi_{\sigma}^+=\sum_{\sigma}\eta_{\sigma}\eta_{\sigma}^+={\bf 1}_{2\times 2}\,.
\end{equation}
The form of these spinors depends on the direction of the spin projection which can be chosen in many ways. In Ref. \cite{CD1} we considered the Pauli spinors of the helicity basis but here we use the spin basis presented in Appendix A with polarizations along the third axis of the rest frame taking ${\bf n}={\bf e}_3$ in Eqs. (\ref{spin1}) and (\ref{spin2}).

It is worth pointing out that the helicity is related to the eigenvalues of the Pauli-Lunabski operator (\ref{PaLu}) while for the polarization $\sigma$ we do not have a differential operator since this is defined in the rest frame. Nevertheless, an operator of the spin projection, $S_3$, can be defined in QFT giving its spectral representation. For this reason we consider here that  the momentum-spin rep. is given by the operators $\{{\bf P},S_3\}$ whose common eigenspinors (\ref{Ups}) and (\ref{Vps}) depend on the eigenvalues $\{{\bf p},\sigma\}$.  However, this set of operators is not complete such that there remain some undetermined integration constants.

The t.m.f. $u_p^{\pm}(t)$ and $v_p^{\pm}(t)$  can be derived   by substituting Eqs. (\ref{Ups}) and (\ref{Vps}) in the Dirac equation (\ref{ED1}).  Then, after a few manipulation, we find the systems of the first order differential equations
\begin{eqnarray}
a(t)\left(i\partial_t\mp m\right)u_p^{\pm}(t)&=&{p}\,u_p^{\mp}(t)\,,\label{sy1}\\
a(t)\left(i\partial_t \mp m\right)v_p^{\pm}(t)&=&-{p}\,v_p^{\mp}(t)\,,\label{sy2}
\end{eqnarray}
in the chart with the proper time or the equivalent system in the conformal chart,
\begin{eqnarray}
\left[i\partial_{t_c}\mp m\, a(t_c)\right]u_p^{\pm}(t_c)&=&{p}\,u_p^{\mp}(t_c)\,,\label{sy1c}\\
\left[i\partial_{t_c} \mp m\, a(t_c)\right]v_p^{\pm}(t_c)&=&-{p}\,v_p^{\mp}(t_c)\,,\label{sy2c}
\end{eqnarray}
which govern the time modulation of the free Dirac field on any spatially flat FLRW manifold. 

The solutions of these systems depend on  integration constants that must be selected according to the charge conjugation (\ref{chc}) which gives the mandatory condition
\begin{equation}\label{VU}
v_p^{\pm}=\left[u_p^{\mp}\right]^*\,.
\end{equation}
The remaining normalization constants can be fixed since the prime integrals of the systems (\ref{sy1}) and (\ref{sy2}), 
$\partial_t (|u_p^+|^2+|u_p^-|^2)=\partial_t (|v_p^+|^2+|v_p^-|^2)=0$, 
allow us to impose the normalization conditions
\begin{equation}
|u_p^+|^2+|u_p^-|^2=|v_p^+|^2+|v_p^-|^2 =1\,, \label{uuvv}\\
\end{equation}
which guarantee that Eqs.  (\ref{ortU}) and (\ref{ortV}) are accomplished. In fact,  we can focus only on the functions $u_p^{\pm}$ since the functions $v_p^{\pm}$ result from Eq. (\ref{VU}). These functions  can be organized as a 2-dimensional space of complex valued vectors $u_p= [ u_p^+, u_p^-]^T$  and $v_p= [ v_p^+, v_p^-]^T$ equipped with the inner product 
\begin{equation}\label{innuv}
(u_p,u'_p)=(v_p,v'_p)= (u_p^{+})^*u_p^{\prime\,+}+(u_p^{-})^*u_p^{\prime\,-}\,.
\end{equation}
Two sets of t.m.f.  $u_p^{a\,\pm},\, a=1,2 $, solutions of the  systems (\ref{sy1}) or (\ref{sy1c}),  which satisfy 
\begin{equation}\label{ortuab}
(u_p^{a}, u_p^{b})=\delta_{ab}\quad, a,b=1,2\,,
\end{equation}
are orthonormal generating a system of  orthonormal fundamental spinors  (\ref{Ups}) and (\ref{Vps}).  Any  linear combination 
\begin{equation}\label{uucc}
u_p^{\pm}=c_1 u_p^{1\,\pm}+c_2 u_p^{2\,\pm}\,,\quad c_1,\,c_2 \in {\Bbb C}\,,
\end{equation}
give rise to normalized spinors  only if 
\begin{equation}\label{cccc}
|c_1|^2+|c_2|^2=1\,.
\end{equation}
Thus we can control the orthogonality of the fundamental spinors exclusively at the level of the  t.m.f., without calculating scalar products.

A special case is that of the rest frame where  the Dirac equation in momentum-spin rep. for  ${\bf p}=0$ can be solved analytically carrying out the normalized fundamental spinors of the rest frame,
\begin{eqnarray}
U_{0,\sigma}(t,{\bf x})&=&\frac{e^{-i mt}}{[2\pi a(t)]^{\frac{3}{2}}}\left(
\begin{array}{c}
\xi_{\sigma}\\
0
\end{array}\right)\,,\label{Ur}\\
V_{0,\sigma}(t,{\bf x})&=&\frac{e^{i mt}}{[2\pi a(t)]^{\frac{3}{2}}}\left(
\begin{array}{c}
0\\
\eta_{\sigma}
\end{array}\right)\,.\label{Vr}
\end{eqnarray} 
However, in the rest frames the energy operator of this picture (\ref{hnp}) takes the simple form  $H=i\partial_t$ since $D\to 0$ when the momentum vanishes. Therefore, the rest frame spinors satisfy the eigenvalue problems $H U_{0,\sigma}(t,{\bf x})=E^+(t) U_{0,\sigma}(t,{\bf x}) $ and $H V_{0,\sigma}(t,{\bf x})=E^-(t) V_{0,\sigma}(t,{\bf x}) $
defining the time-dependent rest energies
\begin{equation}
E_0^{\pm}(t)=\pm m -\frac{3i}{2}\frac{\dot{a}(t)}{a(t)}\,,
\end{equation}
whose real parts are just the rest energies $\pm m$ of special relativity while the imaginary terms are due to the evolution of the background.  Thus we generalize to any FLRW manifold the result we obtained for the dS space-time \cite{CGRG}.

Note that the study of the solutions in rest frames can be done only in the momentum-spin rep. since in the momentum-helicity one the helicity is not defined for vanishing momentum.		

\subsection{Spherical P-waves}

The spherical waves are the solutions with spherical symmetry of the Dirac equation (\ref{EPPN}) that can be derived after introducing the spherical space coordinates of  the frame $\{t,r,\theta,\phi;e\}$ of $(M,a)$. This can be done by rewriting the Dirac operator  (\ref{EDNP}) as  \cite{CD2}
\begin{eqnarray}\label{ED1SS}
E_D(t)&=&i\gamma^0\partial_t+\frac{3i}{2}\frac{\dot{a}(t)}{a(t)}\gamma^0
\nonumber\\
&+&\frac{1}{a(t)}\left(i\frac{1}{r^2}(\gamma^i x^i)\left(x^i\partial_i 
+ 1\right)+i\frac{1}{r^2}\gamma^0(\gamma^i x^i){K}\right),~~~~~~~
\end{eqnarray}
where $r=|{\bf x}|$, pointing out  the angular Dirac operator, 
\begin{equation}\label{DSph}
{K}=\gamma^0 \left(2{\bf L}\cdot{\bf S}+1\right)\,,
\end{equation}
which encapsulates the action of all the angular operators allowing us to separate the spherical variables $(r,\theta,\phi)$ associated to ${\bf x}$. The general solution of this equation,  
\begin{eqnarray}
\psi(t,r,\theta,\phi)&=&\psi^{(+)}(t,r,\theta,\phi)+\psi^{(-)}(t,r,\theta,\phi)\nonumber\\
&=&\int_{0}^{\infty}dp\sum_{\kappa_j,m_j}U_{p,\kappa_j,m_j}(t,r,\theta,\phi){a}(p,\kappa_j,m_j)\nonumber\\
&+&\int_{0}^{\infty}dp\sum_{\kappa_j,m_j}V_{p,\kappa_j,m_j}(t,r,\theta,\phi){b}^{*}(p,\kappa_j,m_j)\,,~~~~~~
\end{eqnarray}
where  $U_{p,\kappa_j,m_j}$  are the fundamental solutions of positive frequencies defined as common eigenspinors of the set $\{{\bf P}^2,K,J_3\}$ corresponding to the eigenvalues $\{p^2, -\kappa_j,m_j\}$ where $\kappa_j=\pm (j+\frac{1}{2})$  \cite{TH}. The eigenspinors of negative frequencies, 
\begin{equation}\label{chUVS}
V_{p,\kappa_j,m_j}(t,r,\theta,\phi)=i\gamma^2 U_{p,\kappa_j,m_j}(t,r,\theta,\phi)^*\,,
\end{equation}
are defined with the help of the charge conjugation as in the case of the plane waves.  Al these spinors may be organized as an orthonormal basis satisfying,
\begin{eqnarray}
\langle U_{p,\kappa_j,m_j}, U_{p',\kappa'_j,m'_j}\rangle &=&
\langle V_{p,\kappa_j,m_j}, V_{p',\kappa'_j,m'_j}\rangle\nonumber\\
&=&\delta_{\kappa_j,\kappa_j^{\prime}}\delta_{m_j,m_j'}\delta(p-p')\,,\label{ortUS}\\
\langle U_{p,\kappa_j,m_j}, V_{p',\kappa'_j,m'_j}\rangle &=&
\langle V_{p,\kappa_j,m_j}, U_{p',\kappa'_j,m'_j}\rangle =0\,, ~~\label{ortVS}
\end{eqnarray}  
with respect to the relativistic scalar product (\ref{SPD1}) that now reads 
\begin{equation}\label{psS}
\left<\psi, \psi'\right>=\int r^2 dr\int_{S^2}d\Omega \, a(t)^3\,\overline{\psi}(t,r,\theta,\phi)\gamma^0\psi'(t,r,\theta,\phi)\,,
\end{equation}
where $d\Omega=d(\cos \theta) d\phi$ is the measure of integration on the sphere $S^2$. These fundamental spinors form the basis of the rep. given by the set of operators $\{{\bf P}^2,K,J_3\}$ which is not complete leaving undetermined some integration constants.

For solving the above eigenvalue problems it is convenient to separate the spherical variables   looking for particular solutions of positive frequencies of the form
\begin{equation}
U_{p,\kappa_j, m_j}(x)=\frac{1}{a(t)^{\frac{3}{2}}}\frac{1}{r}\left[f^+_{p,\kappa_j}(t,r)
\Phi^+_{\kappa_j,m_j}(\theta,\phi)+f^-_{p,\kappa_j}(t,r)\Phi^-_{\kappa_j,m_j}(\theta,\phi)\right]\,,
\end{equation}
where $\Phi^{\pm}_{\kappa_j,m_j}$ are the orthonormal Dirac spherical spinors of special relativity that solve the eigenvalue problems of the operators ${\bf J}^2$, $K$ and $J_3$  as presented in  Appendix B.  Then, after a little calculation by using the identities (\ref{identS}) we derive the system
\begin{equation}\label{rad}
a(t)\left(i\partial_{t}\mp m
\right)f^{\pm}_{p,\kappa_j}(t,r)= \left( \mp\partial_r +
\frac{\kappa_j}{r}\right)f^{\mp}_{p,\kappa_j}(t,r)\,,
\end{equation}
resulted from the Dirac equation (\ref{ED1SS}). In addition,  the eigenvalue problem of
${{\bf P}\,}^2$ leads to the supplemental radial equations
\begin{equation}\label{PP}
\left[-\partial_r^2+\frac{\kappa_j(\kappa_j\pm
1)}{r^2}\right]\rho^{\pm}_{p,\kappa_j}(t,r)=p^2
\rho^{\pm}_{p,\kappa_j}(t,r)\,,
\end{equation}
since the spinors $\Phi^{\pm}_{\kappa_j,m_j}$ are eigenfunctions of
${\bf L}^2$ corresponding to the eigenvalues $\kappa_j(\kappa_j\pm
1)$. 

Under such circumstances, Eqs. (\ref{rad}) and (\ref{PP}) can be solved separating the
variables as,
\begin{equation}
f^{\pm}_{p,\kappa_j}(t,r)=u^{\pm}_p(t)
\rho_{p,\kappa_j}^{\pm}(r)\,,
\end{equation}
finding  that the new functions satisfy \cite{CD2}
\begin{eqnarray}
a(t)\left( i\partial_{t}\mp m
\right)u^{\pm}_p(t)&=&
p \,u^{\mp}_p(t)\,,\label{tmod}\\
\left(\pm \partial_{r}+\frac{\kappa_j}{r} \right)\rho^{\pm}_{p,\kappa_j}(r)&=& p
\,\rho^{\mp}_{p,\kappa_j}(r)\,.\label{rad1}
\end{eqnarray}
Hereby it results that the t.m.f. $u^{\pm}_p$ are the same as in the case of the plane waves satisfying similar equations. Moreover, we assume that these are related as in Eq. (\ref{VU}) and satisfy the normalization conditions  (\ref{uuvv}). Thus the spherical waves have the same t.m.f. as the plane ones as was expected since these have different space shapes but the same time evolution. 

The radial equations (\ref{rad1}) are independent on this time evolution such that we can solve them in terms of Bessel functions. There are two particular solutions 
\begin{eqnarray}
\rho^{1\,\pm}_{p,\kappa_j}(r)&=&\sqrt{pr}J_{\kappa_j\pm \frac{1}{2}}(pr)\,,\\
\rho^{2\,\pm}_{p,\kappa_j}(r)&=&\pm \sqrt{pr}J_{-\kappa_j\mp \frac{1}{2}}(pr)
\end{eqnarray}
which satisfy \cite{LL}
\begin{equation}\label{radnorm}
\int_{0}^{\infty}\rho^{a\,\pm}_{p,\kappa_j}(pr) \rho^{a\,\pm}_{p',\kappa_j}(p'r)dr=\delta(p-p')\,,~~~a=1,2\,,
\end{equation}
thanks to the normalization factor $\sqrt{p}$ introduced above. The general solution, 
\begin{equation}
\rho^{\pm}_{p,\kappa_j}=\hat c_1\rho^{1\,\pm}_{p,\kappa_j}+\hat c_2\rho^{2\,\pm}_{p,\kappa_j}
\end{equation}
keeps this property only if the new integration constants obey $|\hat c_1|^2+|\hat c_2|^2=1$.
We get thus a new integration constant that must be fixed by using supplemental criteria. For example, if we look for solution regular in origin then we must take 
\begin{equation}\label{reg0}
\hat c_1=\frac{1+{\rm sign}\,\kappa_j}{2}\,,\quad
\hat c_2=\frac{1-{\rm sign}\,\kappa_j}{2}\,,
\end{equation}
since the Bessel functions behave as in Eq. (\ref{beh}). Note that this radial solution is more general that in Ref. \cite{CD2} where we adopted the particular version of Ref. \cite{SH}.

Finally, by gathering all the above results and taking into account that the charge conjugation acts on the Dirac spherical spinors as in Eq. (\ref{chsph}), we may write the definitive form of the fundamental spinors 
\begin{eqnarray}
U_{p,\kappa_j, m_j}(x)&=&\frac{1}{a(t)^{\frac{3}{2}}}\frac{1}{r}\left[u^+_p(t) \rho^+_{p,\kappa_j}(r) \Phi^+_{\kappa_j,m_j}(\theta,\phi)\right.\nonumber\\
&&\hspace*{16mm}+\left. u^-_p(t)\rho^-_{p,\kappa_j}(r)\Phi^-_{\kappa_j,m_j}(\theta,\phi)\right]\,,\label{Usph}\\
V_{p,\kappa_j, m_j}(x)&=&\frac{(-1)^{m_j}}{a(t)^{\frac{3}{2}}}\frac{1}{r}\left[-v^+_p(t)\rho^-_{p,-\kappa_j}(r) \Phi^+_{-\kappa_j,-m_j}(\theta,\phi)\right. \nonumber\\
&&\hspace*{18mm}\left.+v^-_p(t)\rho^+_{p,-\kappa_j}(r)\Phi^-_{-\kappa_j,-m_j}(\theta,\phi)\right]\,,~~\label{Vsph}
\end{eqnarray}
which satisfy the charge conjugation symmetry (\ref{chUVS}) and the orthonormalization rules (\ref{ortUS}) and (\ref{ortVS}) if the t.m.f. comply with Eqs. (\ref{VU}) and (\ref{uuvv}).  Moreover, if we have two sets of orthonormal t.m.f. in the sense of Eq. (\ref{ortuab})  then we may construct new solutions as linear combinations in similar conditions as in the case of the plane waves.

\subsection{Adiabatic and rest frame vacua}

The t.m.f.  $u_p^{\pm}(t)$ or $u_p^{\pm}(t_c)$,  are solutions of  the systems (\ref{sy1}) or (\ref{sy1c}) satisfying the conditions  (\ref{VU}) and  (\ref{uuvv}). Unfortunately,  these  are not enough for determining completely these functions such that a supplemental physical hypothesis is required. This is just the criterion of separating the positive and negative frequencies defining thus the vacuum in the momentum reps..    

The vacuum usually considered in Dirac theories is the traditional adiabatic vacuum (a.v.) of the Bunch-Davies type similar to that intensively studied in the case of the scalar fields \cite{BuD}. This can be defined for any FLRW manifold for which the scale factor has the asymptotic behaviour  
\begin{equation}\label{asc}
\lim_{t_c\to-\infty}a(t_c)=0\,.
\end{equation}
Then the asymptotic form of the system (\ref{sy1c}), 
\begin{equation}
i\partial_{t_c}u_p^{\pm}(t_c)=pu_p^{\mp}(t_c)\,,
\end{equation}
gives the behaviour of the modulation functions, 
\begin{equation}
u_p^{\pm}(t_c)\sim c_1 e^{-ipt_c}\pm c_2 e^{ipt_c}\,,
\end{equation}
for $t_c\to -\infty$. According to the common definition,  the a.v. is set when $c_2=0$  since then the modulation functions, $u_p^{+}(t_c)=u_p^{-}(t_c)$, describe a massless particle assumed to be of genuine positive frequency. Thus the general condition of selecting the a.v.  of the Dirac field on FLRW space-times takes the simple form \cite{CrfvD}
\begin{equation}\label{adi}
u_p^{-}(t_c, m)=u_p^{+}(t_c,-m)
\end{equation}
and similarly for the functions $v_p^{\pm}(t_c)$.

The major difficulty of the a.v. as defined above is that in the momentum-spin rep. we cannot reach the rest frame limit.  Indeed, for $p\to 0$ the condition (\ref{adi}) gives the normalized functions
\begin{equation}
\lim_{p\to 0} u_p^{+}(t)=\frac{1}{\sqrt{2}}\,e^{-imt}\,, \quad \lim_{p\to 0}  u_p^{-}(t)=\frac{1}{\sqrt{2}}\,e^{imt}\,,
\end{equation}  
while the limit of $ {\bf \sigma}\cdot {\bf p}$ remains undetermined. Moreover, if we force this limit to zero we obtain  different normalization factors \cite{CQED} such that the limits of the fundamental spinors will  differ from the correct rest spinors (\ref{Ur}) and (\ref{Vr}), mixing thus positive and negative frequencies.  Another impediment is that the a.v. cannot be defined for manifolds whose scale factors do not have a suitable asymptotic behaviour (\ref{asc}).

The solution was to define a new vacuum able to separate the frequencies in any the rest frame of the momentum-spin rep. imposing the conditions
\begin{eqnarray}
\lim_{{\bf p}\to 0} U_{{\bf p},\sigma}(t,{\bf x})&=&U_{0,\sigma}(t,{\bf x})\,,\label{Urf}\\
\lim_{{\bf p}\to 0} V_{{\bf p},\sigma}(t,{\bf x})&=&V_{0,\sigma}(t,{\bf x})\,,\label{Vrf}
\end{eqnarray}  
according to Eqs. (\ref{Ur}) and (\ref{Vr}). These are accomplished if we require the normalized t.m.f. to satisfy \cite{CrfvD}
\begin{equation}\label{rfv}
\lim_{p\to 0} u^{-}_p(t)=\lim_{p\to 0} v^{+}_p(t)=0\,,
\end{equation}
since then the contribution of the matrix $ {\bf \sigma}\cdot {\bf p}$ is eliminated. We say that these conditions define the {\em rest frame vacuum} (r.f.v.) which, in general, is different from the a.v. apart from the Minkowski case in which the t.m.f. 
\begin{equation}
u_p^{\pm}(t)=v_p^{\mp}(t)^*=\sqrt{\frac{E(p)\pm m}{2 E(p)}}\,e^{-i E(p) t}\,,\label{UVMin}
\end{equation}
depending on the energy  $E(p)=\sqrt{p^2 +m^2}$, satisfy simultaneously both the conditions (\ref{adi}) and (\ref{rfv}).

The above definitions of the a.v. and r.f.v. cannot be applied directly to the spherical waves even though these have the same t.m.f.. This is because we have, in addition, the integration constants of the radial functions which have to be fixed. In the case of the a.v. there are no restrictions upon the radial functions such that we can take the convenient t.m.f. according to Eq. (\ref{adi}) but any constants $\hat c_1$ and $\hat c_2$. This is an example in which  the frequency separation is not enough for determining all the integration constants. In this case a good choice is as in Eq. (\ref{reg0}) since then  the radial functions are regular in $r=0$  

In contrast,  the restrictions imposed by the r.f.v. are more effective. This is because in the rest frame, for $p\to 0$, the angular momentum vanishes, ${\bf L}=0$, and, consequently, $K\to \gamma^0$ which means that $\kappa_j=-1$ (corresponding to $l=0$). On the other hand, in this limit the radial function $\rho^{+}_{0,-1}$ is a constant which must remain finite while the other one, $\rho^{-}_{0,-1}$,  is eliminated if we impose the condition (\ref{rfv}).  Taking into account that the Bessel functions behave as in Eq. (\ref{beh}) we draw the conclusion that the only possible choice is 
\begin{equation}\label{ccsph}
\hat c_1=1\,, \quad  \hat c_2=0\,.
\end{equation}
Thus by setting the r.f.v. we determine all the integration constants.

\subsection{Example I: Milne-type universe}   

Let us consider the simple example we proposed recently of a $(1+3)$-dimensional spatially flat FLRW manifold $(M,a)$ with the Milne-type scale factor $a(t)=\omega t$ \cite{Milne} defined for   $t\in (0,\infty)$ such that the conformal time,
\begin{equation}\label{tt}
t_c=\int \frac{dt}{a(t)}=\frac{1}{\omega} \ln(\omega t) ~\to~  a(t_c)\equiv a[t(t_c)]=e^{\omega t_c}\,,
\end{equation}
can take any real value,  $t_c\in (-\infty,\infty)$, and $a(t_c)$ satisfies the condition (\ref{asc}). Note that the free parameter $\omega$, was introduced from dimensional considerations.  We remind the reader that in the case of the genuine Milne's universe (of negative space curvature but globally flat) one must set $\omega=1$ for eliminating the gravitational sources \cite{BD}. 

In contrast, our space-time $(M,a)$ is produced by isotropic gravitational sources, i. e. the density $\rho$ and pressure $p$, evolving in time as
\begin{equation}
\rho=\frac{3}{8\pi G}\frac{1}{t^2}\,, \quad p=-\frac{1}{8\pi G}\frac{1}{t^2}\,,
\end{equation}
and vanishing for $t\to\infty$. These sources govern the expansion of $M$ that can be better observed in the chart $\{t, {\bf \underline x}\}$, of  'observed' space coordinates $\underline{ x}^i=\omega t x^i$, where the line element 
 \begin{equation}
 ds^2=\left(1-\frac{1}{t^2}{\bf \underline x}\cdot {\bf \underline x}\right)dt^2 + 2 {\bf \underline x}\cdot d{\bf \underline x}\,\frac{dt}{t}-d{\bf \underline x}\cdot d{\bf \underline x}\,,
 \end{equation}
lays out an expanding horizon at $|{\bf \underline x}|=t$ and tends to the Minkowski space-time when $t\to \infty$ and the gravitational sources vanish.

On this manifold, we chose the frame  $\{t,{\bf x};e\}$ where  the system (\ref{sy1}) can be analytically solved finding two particular  solutions  \cite{Milne}
\begin{eqnarray}
u_p^{1\,\pm}(t)&=&\sqrt{\frac{m t}{2\pi}} \left[K_{\nu_+(p)}(i m t)\pm K_{\nu_-(p)}(imt)\right]\,,\\
u_p^{2\,\pm}(t)&=&\sqrt{\frac{m t}{2\pi}}  \left[K_{\nu_+(p)}(-i m t)\mp K_{\nu_-(p)}(-imt)\right],~~~~~
\end{eqnarray}
expressed in terms of the  modified Bessel functions,  $K_{\nu_{\pm}(p)}$, of the orders $\nu_{\pm}(p)=\frac{1}{2}\pm i \frac{p}{\omega}$, presented in  Appendix C.  These solutions satisfy Eqs. (\ref{ortuab})  as it results from the  identity (\ref{H3}) with  $\mu=\frac{p}{\omega}$. Therefore, they are orthonormal such the general solution 
\begin{equation}\label{uvcc}
u_p^{\pm}=(v^{\mp}_p)^*=c_1 u_p^{1\,\pm}+c_2 u_p^{2\,\pm}\,,
\end{equation}
is normalized only if the integration constants satisfy the condition (\ref{cccc}).   

In the conformal chart where $a(t_c)=e^{\omega t_c}$ satisfies the asymptotic condition (\ref{asc}),  we can introduce the a.v. imposing the condition (\ref{adi}) which yields  $c_1=c_2=\frac{1}{\sqrt{2}}$. The r.f.v. is given by $c_1=1$ and $c_2=0$ since in the rest frame only the t.m.f. $u_p^{1\,-}$ satisfies the condition (\ref{rfv}) vanishing in the rest linit.  Thus the the t.m.f. of the a.v. and r.f.v. are defined for both the types of solutions, plane and spherical waves. In addition, for the r.f.v. of the spherical waves we must set the radial integration constants as in Eq. (\ref{ccsph}) while in the a.v. these remain arbitrary.

It is worth pointing out that in the  chiral rep. of the Dirac matrices (with diagonal $\gamma^5$) the fundamental spinors of the plane waves in r.f.v. take the simple form \cite{Milne}
\begin{eqnarray}
U_{{\bf p},\sigma}(x)&=&\sqrt{\frac{mt}{\pi}}\frac{e^{i{\bf p}\cdot{\bf x}}}{[2\pi \omega t]^{\frac{3}{2}}}\left(
\begin{array}{c}
K_{\sigma-i\frac{p}{\omega}}\left(im\,t \right) \xi_{\sigma}({\bf p})\\
K_{\sigma+i\frac{p}{\omega}}\left(im\,t \right)\xi_{\sigma}({\bf p})
\end{array}\right)\,,\label{UMin}\\
V_{{\bf p},\sigma}(x)&=&\sqrt{\frac{mt}{\pi}}\frac{e^{-i{\bf p}\cdot{\bf x}}}{[2\pi \omega t]^{\frac{3}{2}}}\left(
\begin{array}{c}
K_{\sigma-i\frac{p}{\omega}}\left(-im\,t \right) \eta_{\sigma}({\bf p})\\
-K_{\sigma+i\frac{p}{\omega}}\left(-im\,t \right)\eta_{\sigma}({\bf p})
\end{array}\right)\,, \nonumber\\\label{VMin}
\end{eqnarray} 
that can be used in applications \cite{Milne}. 

\subsection{Example II: de Sitter expanding universe}

Another example is the well-studied dS expanding universe defined as the expanding portion of the dS manifold where the scale factor has the form (\ref{Hubb}). In the frame $\{t_c,{\bf x};e\}$ of the conformal time, 
\begin{equation}
t_c=-\frac{1}{\omega}e^{-\omega t}\in (-\infty, 0]~~~ \to~~~ a(t_c)=-\frac{1}{\omega t_c}\,.
\end{equation}
we may derive two particular solutions of the system (\ref{sy1c}), 
\begin{eqnarray}
u_p^{1\,\pm}(t_c)&=&\sqrt{-\frac{p t_c}{\pi}} K_{\nu_{\mp}}(ipt_c)\,,\label{dSup}\\
u_p^{2\,\pm}(t_c)&=&\pm \sqrt{-\frac{p t_c}{\pi}} K_{\nu_{\mp}}\left(-i p t_c\right)\,,\label{dSum}
\end{eqnarray}
expressed in terms of modified Bessel functions of the orders $\nu_{\pm}=\frac{1}{2}\pm i \mu$ with $\mu=\frac{m}{\omega}$. According to Eq. (\ref{H3}) we find that these t.m.f. are orthonormal, satisfying Eqs. (\ref{ortuab}), giving rise to the particular spinors 
\begin{equation}\label{partU}
U_{{\bf p},\sigma}^{1/2}(t,{\bf x}\,)=(\omega t_c)^2
\frac{e^{i{\bf p}\cdot{\bf x}}}{(2\pi )^{\frac{3}{2}}}\sqrt{\frac{p}{\pi\omega}}\left(
\begin{array}{c}
K_{\nu_-}(\pm i p t_c) \,
\xi_{\sigma}\\
\pm K_{\nu_+}(\pm i p t_c) \,
{\bf \sigma}\cdot{\bf n}_p\,\xi_{\sigma}
\end{array}\right)\,,
\end{equation}
which are orthonormal. Then the linear combination (\ref{uucc}) gives the general solution
\begin{equation}
U_{{\bf p},\sigma}=c_1U_{{\bf p},\sigma}^{1}+c_2 U_{{\bf p},\sigma}^{2}\,,
\end{equation}
which are  normalized only if the integration constants satisfy the condition (\ref{cccc}). The corresponding spinors of negative frequencies can be obtained from Eq. (\ref{chc}).

The a.v. can be defined simply by choosing $c_1=1$ and $c_2=0$ as in Ref. \cite{CD1} since then the condition (\ref{adi}) is accomplished as we can deduce from  Eq. (\ref{Km0}). This vacuum is different from the rest frame one which must comply with the condition (\ref{rfv}).  Taking into account that the Bessel functions behave as in Eq. (\ref{beh}) we find that now we obtain the constants  
\begin{equation}\label{con}
c_1=\frac{e^{\pi \mu}p^{-i\mu}}{\sqrt{1+e^{2\pi \mu}}}\,,\quad c_2=\frac{i p^{-i\mu}}{\sqrt{1+e^{2\pi \mu}}}\quad \,.
\end{equation}
that  can be seen as the Bogolyubov coefficients of the transformation between the orthonormal bases corresponding to the a.v. and r.f.v. 
Furthermore, by using the connection formula (\ref{KII}), we obtain the definitive form of the t.m.f.  (\ref{uucc})  in the r.f.v. as \cite{CrfvD}
\begin{equation}
u_p^{\pm}(t_c)=\pm \frac{\sqrt{-\pi t_c}\, p^{\nu_-} }{\sqrt{1+e^{2\pi \mu}}}\, I_{\mp\nu_{\mp}}(ipt_c)
\end{equation}
which have the  remarkable property  
\begin{equation}
\lim_{t_c\to 0}|u^{+}_p(t_c)|=1\,, \quad \lim_{t_c\to 0}u^{-}_p(t_c)=0\,,
\end{equation}
that may be interpreted as an adiabatic condition for $t\to \infty$ instead of  $t\to- \infty$. The t.m.f. of the negative frequencies have to be calculated according to Eq.(\ref{VU}). 
For the spherical waves we have to use the same t.m.f. with arbitrary radial integration constants in the a.v. or satisfying the condition (\ref{ccsph}) if we set the r.f.v..

We must specify that these t.m.f. are defined up to an arbitrary phase factor depending on $p$ which may assure the correct flat limits of the plane or spherical waves, determining the  forms of the one particle operators of the QFT \cite{prep}.   

\section{Energy representations in de Sitter expanding universe}

We studied so far the solutions of the Dirac equation that can be obtained separating the Cartesian or spherical variables in NP. All these solutions depend on the vector or scalar momentum which is conserved on any FLRW space-time. However, on the dS expanding universe we have, in addition, a conserved energy operator that in the  SP takes the familiar form  (\ref{hsp}). This suggest us to look for states of given energy that could be derived on this time evolution picture of the dS manifold. 

\subsection{Plane E-waves in SP}

We consider first the plane waves in the dS expanding universe where  from Eqs. (\ref{HXP}) and  (\ref{Hubb}) we deduce the commutation relation $[H,P^i]=i \omega P^i$ showing  that the energy and the  momentum components cannot be measured simultaneously with a desired accuracy \cite{CGRG}. Nevertheless, this commutation relation does not affect the direction of the momentum operator which encourage us to define the new operators of SP, namely  the scalar momentum, $P_S$, and the operators  $N^i$ of the direction of propagation such that $P^i_S=P_S N^i$ and
\begin{equation}
[H_S,P_S]=i P_S\,, \quad [H_S,N^i]=0\,.
\end{equation}
Then we can chose the system of commuting operators $\{H_S,N^i\}$ for defining the energy rep. we look for. The difficulty is that the operators $N^i$ are no longer differential operators such that their eigenvalues have to be pointed out indirectly when we construct the solutions. 

Let us consider now  the Dirac equation (\ref{EPPS})  in the frame $\{t,{\bf x};e\}$ of SP of the dS manifold  whose operator  (\ref{EDSP}) becomes independent on time,
\begin{equation}\label{EDSS}
E_D^S=i\gamma^0\partial_{t}-{\bf \gamma}\cdot{\bf P}
- \omega\gamma^{0}\hat D\,,
\end{equation}
as it results from Eq. (\ref{Hubb}). The first step is to rewrite this equation in momentum rep. starting with the $4$-dimensional Fourier integral rep., 
\begin{eqnarray}
\psi_S(x)&=&\psi^{(+)}_S(x)+\psi^{(-)}_S(x)\nonumber\\
&=&\int_0^{\infty}\,dE\int_{{\Bbb R}^3_p}\, d^3p\,\, \left[\hat{\psi}^{(+)}_S(E,{\bf p})\,e^{-i(Et-{\bf p}\cdot {\bf x})}\right.\nonumber\\
&&\hspace*{33mm}\left. +\,\hat{\psi}^{(-)}_S(E,{\bf p})\,e^{i(Et-{\bf p}\cdot {\bf x})}\right]\,,\label{PsiS}
\end{eqnarray}
where $\hat{\psi}^{(\pm)}_S$ are spinors of positive and respectively negative frequencies. Assuming that these spinors  behave as tempered distributions
on the domain ${\Bbb R}_p^3$ we may apply the Green theorem when the Dirac operator (\ref{EDSS}) acts on $\psi_S(x)$. Then we may replace the momentum operators ${P}^i$ by  ${p}^i$ and the coordinate operators ${X}^i$ by $i {\partial}_{p_i}$  obtaining the general form of the  Dirac equation of SP in momentum rep.,
\begin{equation}\label{ED4}
\left[\pm E\gamma^0\mp{\gamma}^i{p}^i -m -i\gamma^{0}\omega \left({p}^i{
\partial}_{p_i}+\frac{3}{2}\right)\right]\hat{\psi}^{(\pm)}_S(E,{\bf p})=0\,,
\end{equation}
where $E$ is the energy defined as the eigenvalue of the operator $H_S=i\partial_t$. 

Denoting then ${\bf p}=p\, {\bf n}_p$ with $p=|{\bf p}\,|$, we observe that the differential operator of Eq. (\ref{ED4}) is of radial type, ${p}^i{\partial}_{p_i}=p\,\partial_p$. Therefore, this operator acts on the functions which depend on $p$  while the functions which depend only on the momentum direction ${\bf n}_p$  behave as constants. This suggest us to look for solutions of the form \cite{CD4}
\begin{eqnarray}
&&\hat{\psi}^{(+)}_S(E,{\bf p})=\sum_{\sigma}u^S({E,{\bf p},\sigma})\,a(E,{\bf n}_p,\sigma)\,,\\
&&\hat{\psi}^{(-)}_S(E,{\bf p})=\sum_{\sigma}v^S({E,{\bf p},\sigma})\,b^*(E,{\bf n}_p,\sigma)\,,
\end{eqnarray}
where the wave functions  ${a}$ and ${b}$ play the role of constants as long as they do not depend on $p$.

Furthermore, we assume that, in the standard rep. of the Dirac matrices,  the spinors of the momentum rep. complying with the charge conjugation symmetry  have the form
\begin{eqnarray}
u^S(E,{\bf p},\sigma)&=&\left(
\begin{array}{c}
f^{+}_E(p)\,\xi_{\sigma}\\
 f^{-}_E(p)\,{\bf \sigma}\cdot{\bf n}_p\,\xi_{\sigma}
\end{array}\right)\,,\label{uS}\\
v^S(E,{\bf p},\sigma)&=&\left(
\begin{array}{c}
 f^{-}_E(p)^*\,{\bf \sigma}\cdot{\bf n}_p\,\eta_{\sigma}\\
f^{+}_E(p)^*\,\eta_{\sigma}
\end{array}\right)\,,\label{vS}
\end{eqnarray}
where $\xi_{\sigma}$ and $\eta_{\sigma} $ are the Pauli spinors of an arbitrary spin basis. After a little calculation we find that, according to Eq. (\ref{ED4}), the radial functions satisfy the system
\begin{equation}
\left[i\omega \left(p\,\frac{d}{dp}
+\frac{3}{2}\right)-(E\mp m)\right]f^{\pm}_E(p)=- p\, f^{\mp}_E(p)\,,
\end{equation}
that is analytically solvable. Indeed, after denoting  $\mu=\frac{m}{\omega}$ and
$\epsilon=\frac{E}{\omega}$, we find that the general solutions are linear combinations,
\begin{equation}
f^{\pm}_E(p)= c_1 \phi^{\pm}_1(s)+c_2 \phi_2^{\pm}(s)\,,\label{fpus}
\end{equation}
of the particular solutions  \cite{CD4}
\begin{eqnarray}
\phi_{1}^{\pm}(s)&=&N_1 s^{-1-i\epsilon} K_{\nu_{\mp}}(-i s)\,,\label{fiplus1}\\
 \phi_2^{\pm}(s)&=&\pm N_2 s^{-1-i\epsilon} K_{\nu_{\mp}}(i s)\,,\label{fiplus2}
\end{eqnarray}
depending on the new variable
\begin{equation}\label{X2}
s=\frac{p}{\omega}
\end{equation} 
and the indices $\nu_{\pm}=\frac{1}{2}\pm i\mu$. 
The normalization constants $N_{1,2}$ have to assure the normalization in the energy scale for each particular solution separately.

Collecting all the above results we can write down the final expression of the
Dirac field (\ref{PsiS}) as
\begin{eqnarray}
\psi_S(x)&=&\int_0^{\infty}\,dE\int_{S^2}\, d\Omega_n\,\,
\sum_{\sigma}\left[U^S_{E,{\bf n}_p,\sigma}(t,{\bf x}) a(E,{\bf n}_p,\sigma)
\right.\nonumber\\
&&~~~~~~~~~~~~\left.+\, V^S_{E,{\bf n}_p,\sigma}(t,{\bf x}) b^*(E,{\bf n}_p,\sigma)\right],
\end{eqnarray}
where the integration covers the energy semi-axis and  the sphere $S^2$. The notation $U^S$ and $V^S$ stands for the fundamental spinor solutions of positive and,
respectively, negative frequencies of energy $E$, momentum direction ${\bf n}_p$ and polarization $\sigma$. According to Eqs. (\ref{uS}), 
(\ref{fpus}),  (\ref{fiplus1})  and  (\ref{fiplus2}) we can write  the spinors of  positive frequencies  as the linear combination
\begin{equation}\label{Utot}
U^{S}_{E,{\bf n}_p,\sigma}=c_1 U^{S\,1}_{E,{\bf n}_p,\sigma}+c_2 U^{S\,2}_{E,{\bf n}_p,\sigma}\,,
\end{equation}
of the particular spinors defined by the integral reps. \cite{CD4}
\begin{equation}
U^{S\, 1,2}_{E,{\bf n}_p,\sigma}(t,{\bf x})=e^{-iEt} \int_{0}^{\infty} s^2\, ds\,\left(
\begin{array}{c}
\phi^+_{1,2}(s)\, \xi_{\sigma}\\
\phi^-_{1,2}(s)\,{\bf \sigma}\cdot{\bf n}_p\, \xi_{\sigma}
\end{array}\right)
e^{i \omega s {\bf n}_p\cdot{\bf x}}\,,\label{US12}
\end{equation}
while the negative frequencies ones are their charge conjugated spinors, 
\begin{equation}\label{conj}
V^S_{E,{\bf n}_p,\sigma}=(U^S_{E,{\bf n}_p,\sigma})^{c}=i\gamma^2
({U}^S_{E,{\bf n}_p,\sigma})^*\,.
\end{equation}
After integrating over $s=\frac{p}{\omega}$, the unit vector ${\bf n}_p$ giving the direction of propagation  becomes an independent parameter that will be denoted from now simply as ${\bf n}$ considering that its components, $n^i,$ are just the eigenvalues of the operators $N^i$.  Note that in QFT these operators can be defined as one particle operators giving directly their spectral representations.

The functions $f^{\pm}_E$ of Eq. (\ref{fpus}) are defined up to the integration constants $c_1$ and $c_2$ and the normalization factors $N_1$ and $N_2$ which must be fixed in order to assure  the orthonormalization relations
\begin{eqnarray}
&&\left<U^S_{E,{\bf n},\sigma},U^S_{E,{\bf n}^{\,\prime},\sigma^{\prime}}\right>_S=
\left<V^S_{E,{\bf n},\sigma},V^S_{E,{\bf n}^{\,\prime},\sigma^{\prime}}\right>_S\nonumber\\
&&\hspace*{14mm}= \delta_{\sigma\sigma^{\prime}}\delta(E-E')\,\delta^2 ({\bf n}-{\bf n}^{\,\prime})\,,
\label{orto1}\\
&&\left<U^S_{E,{\bf n},\sigma},V^S_{E,{\bf
n}^{\,\prime},\sigma^{\prime}}\right>_S= \left<V^S_{E,{\bf
n},\sigma},U^S_{E,{\bf n}^{\,\prime},\sigma^{\prime}}\right>_S=
0\,,~~~~~~\label{orto2}
\end{eqnarray}
and the completeness condition of the form (\ref{complet}) corresponding to the scalar product (\ref{SPSP}). 
Moreover, we require the particular spinors $U^{S\,1,2}_{E,{\bf n},\sigma}$ to form an orthonormal system. We apply the method of the Appendix D for calculating  the scalar products (\ref{IntUU}) of these particular solutions obtaining  that $U^{S\,1}_{E,{\bf n},\sigma}$ and $U^{S\,2}_{E,{\bf n},\sigma}$ are orthogonal and  the conditions (\ref{orto1}) and (\ref{orto2}) are accomplished if we take  \cite{CD4}
\begin{equation}\label{norm}
N_1=N_2=N=\frac{1}{(2\pi)^{3/2}}\, \frac{\omega}{\sqrt{2}\pi}\,.
\end{equation}
while  the integration constants satisfy
\begin{equation}
|c_1|^2+|c_2|^2=1\,.
\end{equation}
assuring  the correct normalization of the general spinors (\ref{Utot}).  

Hence we derived the most general system of fundamental spinors that form the generalized basis of the energy-spin rep. $\{H_S,N^i,S_3\}$ in which the spinors depend on the eigenvalues $\{E,{\bf n},\sigma\}$.  Obviously, these are determined up to an integration constant which has to be fixed according to supplemental criteria. In Ref. \cite{CD4} we fixed {\em a priori}  for simplicity $c_1=1$ and $c_2=0$ without other arguments. 

The integral rep. of the fundamental spinors is useful for calculating scalar products as in the Appendix D but the definitive form of these spinors may be obtained after performing these integrals. This is missing in Ref. \cite{CD4} such that we perform this calculation for the first time here finding that the particular spinors  (\ref{US12}) can be written as 
\begin{equation}\label{US12X}
U^{S\, 1,2}_{E,{\bf n},\sigma}(t,{\bf x})
=e^{-iEt}\left(
\begin{array}{c}
{\cal F}^+_{1,2}(z)\, \xi_{\sigma}\\
{\cal F}^-_{1,2}(z)\,{\bf \sigma}\cdot{\bf n}\, \xi_{\sigma}
\end{array}\right)
\end{equation}
in terms of the new functions,
\begin{eqnarray}
{\cal F}^{\pm}_1(z)&=&~\frac{N}{2}\,e^{\frac{\pi\epsilon}{2}}\left[A_{\pm}(\epsilon,\mu, z)-B_{\pm}(\epsilon,\mu, z)\right]\,,\\
{\cal F}^{\pm}_2(z)&=&\pm\frac{N}{2}\,e^{-\frac{\pi\epsilon}{2}}\left[A_{\pm}(\epsilon,\mu, z)+B_{\pm}(\epsilon,\mu, z)\right]\,,~~~~
\end{eqnarray}  
depending on the dimensionless variable $z=\omega\, ({\bf n}\cdot{\bf x})$, where the quantities
\begin{eqnarray}
A_{\pm}(\epsilon,\mu,z)&=&\frac{1}{2}\,\Gamma\left(\textstyle{-\frac{1}{4}-\frac{i\epsilon}{2}\pm\frac{i\mu}{2}}\right)\Gamma\left(\textstyle{\frac{1}{4}-\frac{i\epsilon}{2}\mp\frac{i\mu}{2}}\right)\nonumber\\
&\times&  F\left(\textstyle{-\frac{1}{4}-\frac{i\epsilon}{2}\pm\frac{i\mu}{2},\frac{1}{4}-\frac{i\epsilon}{2}\mp\frac{i\mu}{2}};\textstyle{\frac{1}{2}};\,z^2 \right)\,,\label{Apm}\\
B_{\pm}(\epsilon,\mu,z)&=&z\,\Gamma\left(\textstyle{\frac{3}{4}-\frac{i\epsilon}{2}\mp\frac{i\mu}{2}}\right)\Gamma\left(\textstyle{\frac{1}{4}-\frac{i\epsilon}{2}\pm\frac{i\mu}{2}}\right)\nonumber\\
&\times&  F\left(\textstyle{\frac{3}{4}-\frac{i\epsilon}{2}\mp\frac{i\mu}{2},\frac{1}{4}-\frac{i\epsilon}{2}\pm\frac{i\mu}{2};\frac{3}{2}};\,z^2 \right)\,,\label{Bpm}
\end{eqnarray}
have the remarkable properties
\begin{eqnarray}
A_{\pm}(\epsilon,\mu,z)&=& A_{\mp}(\epsilon,-\mu,z)\,,\\
 B_{\pm}(\epsilon,\mu,z)&=&B_{\mp}(\epsilon,-\mu,z)\,,
\end{eqnarray}
that could be of interest in interpreting the above solutions obtained in SP. Then Eq. (\ref{conj}) will give the corresponding  particular spinors $V^{S\, 1,2}_{E,{\bf n},\sigma}$. This closed form of the plane E-waves is obtained here for the first time.

The surprise in deriving these solutions is that, after separating the frequencies by using directly the eigenvalues of the energy operator, we remain with undetermined integration constants since in this case the integration constants are not related to the manner in which the vacuum is defined. 

\subsection{Plane E-waves in NP}

The last step is to rewrite the final results in NP where the Dirac field,
\begin{eqnarray}
\psi(x)&=&T(t)^{-1}\psi_S(t,{\bf x})=\psi_S(t,e^{\omega t}{\bf x})\nonumber\\
&=&\int_0^{\infty}\,dE\int_{S^2}\,
d\Omega_n\,\,
\sum_{\sigma}\left[U_{E,{\bf n},\sigma}(t,{\bf x}) a(E,{\bf n},\sigma)\right.\nonumber\\
&&\hspace*{38mm}+\left. V_{E,{\bf n},\sigma}(t,{\bf x}) b^*(E,{\bf n},\sigma)\right]\,,
\end{eqnarray}
depends on the fundamental spinors of the energy-spin basis of  NP  that may have the general form
\begin{eqnarray}
U_{E,{\bf n},\sigma}&=&c_1 U_{E,{\bf n},\sigma}^1+c_2 U_{E,{\bf n},\sigma}^2\,,\label{Ucc}\\
V_{E,{\bf n},\sigma}&=&c_1^* V_{E,{\bf n},\sigma}^1+c_2^* V_{E,{\bf n},\sigma}^2\,,\label{Vcc}
\end{eqnarray}
depending and the particular spinors 
\begin{eqnarray}
U_{E,{\bf n},\sigma}^{1,2}(t,{\bf x})&=&U^{S\,1,2}_{E,{\bf n},\sigma}(t,e^{\omega t}{\bf x})\nonumber\\
&=&e^{-iEt} \left(
\begin{array}{c}
{\cal F}^{+}_{1,2}(z\,e^{\omega t})\, \xi_{\sigma}\\
{\cal F}^{-}_{1,2}(z\,e^{\omega t})\,{\bf \sigma}\cdot{\bf n}\, \xi_{\sigma}
\end{array}\right)\,,\label{UNPtx}\\
V_{E,{\bf n},\sigma}^{1,2}(t,{\bf x})&=&V^{S\,1,2}_{E,{\bf n},\sigma}(t,e^{\omega t}{\bf x})\nonumber\\
&=&e^{iEt}\left(
\begin{array}{c}
{\cal F}^{-}_{1,2}(z\,e^{\omega t})^*\,{\bf \sigma}\cdot{\bf n}\, \eta_{\sigma}\\
{\cal F}^{+}_{1,2}(z\,e^{\omega t})^*\, \eta_{\sigma}
\end{array}\right)\,,\label{VNPtx}
\end{eqnarray}
which have a more complicated time dependence since in NP the time and space variables are no longer separable. 

The spinors (\ref{Ucc}) and (\ref{Vcc}) are eigenspinors of the energy operator (\ref{hnp})  of NP corresponding to positive and negative frequencies which are separated as in special relativity according to the sign of the energy. For each pair of constants $c_1$ and $c_2$ satisfying the condition (\ref{cccc}) they form a basis of the rep. $\{H,K,J_3\}$ of NP but these constants remain unspecified playing the role of free parameters. Note that the criteria  that  fixed the a.v. or r.f.v. do not hold here since these conditions depend on momentum which is not defined in this rep.. We remain thus with an ambiguity that could be helpful in further developments.

The closed forms of the particular solutions (\ref{UNPtx}) and (\ref{VNPtx}) are not suitable for concrete calculations. For this reason we turn back to the integral representations that could offer us more flexibility. Let us focus on the particular solutions in the form  (\ref{US12}) that in the frame $\{t_c,{\bf x};e\}$ of NP can be rewritten as
\begin{eqnarray}
&&U^{1/2}_{E,{\bf n},\sigma}(t_c,{\bf x})=U^{S\,1/2}_{E,{\bf n},\sigma}(t,{\bf x}\,e^{\omega t})\nonumber\\
&&~~~~~~=N (\omega t_c)^2 \int_{0}^{\infty}  ds\, s^{1-i\epsilon}\left(
\begin{array}{c}
K_{\nu_-}(\pm s\omega t_c)\, \xi_{\sigma}\\
\pm K_{\nu_+}(\pm s\omega t_c)\,{\bf \sigma}\cdot{\bf n}\, \xi_{\sigma}
\end{array}\right)
e^{i \omega s {\bf n}\cdot{\bf x}}\,,~~~~~~~\label{US12NP}
\end{eqnarray}
after changing the integration variable as $s\to s e^{-\omega t}=-s\, \omega t_c$. With these integral representations we can calculate scalar products in NP.

We verify first that the spinors (\ref{UNPtx}) and (\ref{VNPtx})  are orthonormal with respect to the scalar product (\ref{SPD1}) of NP. For those of positive frequency we obtain
\begin{equation}
\left<U_{E,{\bf n},\sigma}^a,U_{E',{\bf n}^{\,\prime},\sigma'}^b\right>=\delta_{ab}\delta(E-E')\delta^2({\bf n}-{\bf n}^{\,\prime})\,, 
\end{equation}
and similarly for the negative frequencies ones. Furthermore,  we calculate the transition coefficients between the basis of the momentum and energy reps. of NP obtaining that the above particular spinors and those of the momentum-spin rep. (\ref{partU}) satisfy, 
\begin{eqnarray}
&&\left<U_{{\bf p},\sigma}^a,U_{E,{\bf n},\sigma'}^b\right>= \delta_{ab}\delta_{\sigma,\sigma'}\delta^2({\bf n}-{\bf n}_p)\nonumber\\
&&~~~~~~~~\times  \frac{1}{\sqrt{2\pi}\,\omega^2}\left(\frac{p}{\omega}\right)^{-\frac{3}{2}-i\frac{E}{\omega}}\,, ~~ a,b=1,2\,, 
\end{eqnarray}
pointing our  the isometry between the bases of P and E-plane waves having the same integration constants $c_1$ and $c_2$. Moreover, by using the inversion relations (\ref{invers}) we can relate the particle wave functions of these reps. in the dS space-time  as
\begin{eqnarray}
a({\bf p},\sigma)&=&\int_0^{\infty}dE\int_{S^2}d\Omega_n \sum_{\sigma '}\langle U_{{\bf p},\sigma},U_{E,{\bf n},\sigma'}\rangle a(E,{\bf n},\sigma')\nonumber\\
&=&\frac{p^{-3/2}}{\sqrt{2\pi\omega}}\int_0^{\infty}dE\,
\left(\frac{p}{\omega}\right)^{-i\frac{E}{\omega}}\,a(E,{\bf n}_p,\sigma)\,,\label{Iaa1}\\
a(E,{\bf n},\sigma)&=&\int d^3 p \,\sum_{\sigma'}\langle U_{E,{\bf n},\sigma},U_{{\bf p},\sigma'}\rangle a({\bf p},\sigma')\nonumber\\
&=&\frac{1}{\sqrt{2\pi\omega}}\int_0^{\infty}dp\,\sqrt{p}\,\,
\left(\frac{p}{\omega}\right)^{i\frac{E}{\omega}}\,a(p {\bf n},\sigma)\,,\label{Iaa2}
\end{eqnarray}
and similarly for the anti-particle wave functions $b$. It is remarkable that these relations are very similar to those we found previously for the scalar \cite{KG} and Maxwell \cite{Max} fields.  

Finally we note that this isometry does not solve the problem of the undetermined  integration constants even though now we can to take over the constants of P-waves for obtaining an isometric basis of E-waves. This is because the criteria used for defining the vacua of the P-waves become meaningless  for the E-waves where the frequencies are already separated.   In our opinion the problem of fixing these constants remains open. 

\subsection{Spherical E-waves}

The solutions we present here as a premier are of the energy-angular momentum rep. that solve the Dirac equation (\ref{EPPS})  in the frame $\{t, r,\theta,\phi;e\}$ of SP of the dS manifold.  In this frame the time-independent Dirac operator (\ref{EDSS})  can be rewritten as   
\begin{eqnarray}\label{ED1S}
E_D^S&=&i\gamma^0\partial_t+i\gamma^0 \omega\left(x^i\partial_i+\frac{3}{2}\right)\nonumber\\
&+&i\frac{1}{r^2}(\gamma^i x^i)\left(x^i\partial_i + 1\right)
+\,i\frac{1}{r^2}\gamma^0(\gamma^i x^i){K} \,,~~~~~
\end{eqnarray}
keeping the notation $r=|{\bf x}|$ and $K$ for the spherical Dirac operator (\ref{DSph}).

We look for  for general solutions of the form 
\begin{eqnarray}
\psi_S(t,r,\theta,\phi)&=&\psi_S^{(+)}(t,r,\theta,\phi)+\psi_S^{(-)}(t,r,\theta,\phi)\nonumber\\
&=&\int_{0}^{\infty}dE\sum_{\kappa_j,m_j}U^S_{E,\kappa_j,m_j}(t,r,\theta,\phi)a(E,\kappa_j,m_j)\nonumber\\
&+&\int_{0}^{\infty}dE\sum_{\kappa_j,m_j}V^S_{E,\kappa_j,m_j}(t,r,\theta,\phi){b}^{*}(E,\kappa_j,m_j)\,,~~~~~
\end{eqnarray}
where  $U_{E,\kappa_j,m_j}$  are the fundamental solutions of positive frequencies defined as common eigenspinors of the set $\{ H_S,K,J_3\}$ corresponding to the eigenvalues $\{E, -\kappa_j,m_j\}$ where the energy $E$ is the eigenvalue of the energy operator $H_S=i\partial_t$ of this picture.  The eigenspinors of negative frequencies, 
\begin{equation}\label{chUVS1}
V^S_{E,\kappa_j,m_j}(t,r,\theta,\phi)=i\gamma^2 U^S_{E,\kappa_j,m_j}(t,r,\theta,\phi)^*\,,
\end{equation}
are defined with the help of the charge conjugation as in the case of the plane waves.  Al these spinors may be organized as the orthonormal angular momentum basis satisfying,
\begin{eqnarray}
\langle U^S_{E,\kappa_j,m_j}, U^S_{E',\kappa'_j,m'_j}\rangle_S &=&
\langle V^S_{E,\kappa_j,m_j}, V^S_{E',\kappa'_j,m'_j}\rangle_S\nonumber\\
&=&\delta_{\kappa_j,\kappa_j^{\prime}}\delta_{m_j,m_j'}\delta(E-E')\,,~~~~\label{ortUS1}\\
\langle U^S_{E,\kappa_j,m_j}, V^S_{E',\kappa'_j,m'_j}\rangle_S &=&
\langle V^S_{E,\kappa_j,m_j}, U^S_{E',\kappa'_j,m'_j}\rangle_S\nonumber\\
 &=&0\,, \label{ortVS1}
\end{eqnarray}  
with respect to the relativistic scalar product (\ref{SPSP}) that now reads 
\begin{equation}\label{psS1}
\left<\psi, \psi'\right>_S=\int r^2 dr\int_{S^2}d\Omega \, \overline{\psi}_S(t,r,\theta,\phi)\gamma^0\psi_S'(t,r,\theta,\phi)\,,
\end{equation}
where we integrate on SPhere $S^2$ as in Eq. (\ref{psS}).

For solving the above eigenvalue problems it is convenient to separate the time and the spherical variables  looking for particular solutions of positive frequencies of the form
\begin{equation}
U^S_{E,\kappa_j, m_j}(x)=\frac{e^{-iEt}}{r}\left[\rho^+_{E,\kappa_j}(r)
\Phi^+_{\kappa_j,m_j}(\theta,\phi)+\rho^-_{E,\kappa_j}(r)\Phi^-_{\kappa_j,m_j}(\theta,\phi)\right]\,,
\end{equation}
where $\Phi^{\pm}_{\kappa_j,m_j}$ are the orthonormal Dirac spherical spinors of  the Appendix B.  Then, after a little calculation by using the identities (\ref{identS}) we derive the system
\begin{equation}\label{rad2}
\left( E\pm m + i \omega r\frac{d}{dr} + \frac{i\omega}{2}
\right)\rho^{\pm}_{E,\kappa_j}= \left(\mp\frac{d}{dr}+
\frac{\kappa_j}{r}\right)\rho^{\mp}_{E,\kappa_j}\,,
\end{equation}
resulted from the Dirac equation (\ref{ED1S}) which can be rewritten as
\begin{eqnarray}
(1-\omega^2 r^2) \frac{d\rho^+_{E,\kappa_j}}{dr}-\left(2\omega^2 \beta r+\omega^2 \kappa_j r-\frac{\kappa_j}{r} \right) \rho^+_{E,\kappa_j}
&=&2i\omega\alpha \rho^-_{E,\kappa_j}\,,\label{E1}\\
(1-\omega^2 r^2) \frac{d\rho^-_{E,\kappa_j}}{dr}-\left(2\omega^2 \alpha r-\omega^2 \kappa_j r+\frac{\kappa_j}{r} \right) \rho^-_{E,\kappa_j}
&=&-2i\omega\beta \rho^+_{E,\kappa_j}\,,~~~~\label{E2}
\end{eqnarray}
keeping the previous notations, $\mu=\frac{m}{\omega}$ and $\epsilon=\frac{E}{\omega}$, and defining the parameters
\begin{eqnarray}
\alpha&=&\frac{1}{4}+\frac{\kappa_j}{2}-\frac{i\epsilon}{2}-\frac{i\mu}{2}\,,  \\
\beta&=&\frac{1}{4}-\frac{\kappa_j}{2}-\frac{i\epsilon}{2}+\frac{i\mu}{2}\,. 
\end{eqnarray}
In this manner, after separating the time and angular variables, we remain with a radial problem in the spaces of the 2-dimensional vectors ${\cal R}_{E,\kappa_j}=\left[\rho^+_{E,\kappa_j},\rho^-_{E,\kappa_j} \right]$ which must satisfy the radial orthonormalization condition
\begin{equation}
\left<{\cal R}_{E,\kappa_j},{\cal R}_{E',\kappa_j}\right>=\int_0^{\infty} dr {\cal R}_{E,\kappa_j}(r){\cal R}_{E',\kappa_j}^{\,+}(r)
=\delta(E-E')\,,\label{norad}
\end{equation}
resulted from  Eqs. (\ref{ortUS1}), (\ref{psS1}) and the orthogonality of the spherical spinors.  The system of Eqs. (\ref{E1}) and (\ref{E2}) can be solved analytically in terms of Gauss hypergeaometric functions obtaining two particular solutions of the form 
\begin{eqnarray}
&&{\cal R}_{E,\kappa_j}^1(r)^T=N_1 (\omega r)^{-\kappa_j}\nonumber\\
&&~~~~\times \left[
\begin{array}{c}
F\left(\alpha-\kappa_j+\frac{1}{2},\beta;\frac{1}{2}-\kappa_j;\omega^2 r^2 \right)\\
\frac{\textstyle2i\beta }{\textstyle2\kappa_j-1}\,\omega r F\left(\alpha-\kappa_j+\frac{1}{2},\beta+1;\frac{3}{2}-\kappa_j;\omega^2 r^2 \right)
\end{array}\right]\,,\\
&&{\cal R}_{E,\kappa_j}^2(r)^T=N_2 (\omega r)^{\kappa_j}\nonumber\\
&&~~~~\times \left[
\begin{array}{c}
\frac{\textstyle 2i\alpha}{\textstyle 2\kappa_j+1}\,\omega r F\left(\alpha+1,\beta+\kappa_j+\frac{1}{2},\beta;\frac{3}{2}+\kappa_j;\omega^2 r^2 \right)\\
 F\left(\alpha, \beta+\kappa_j+\frac{1}{2};\frac{1}{2}+\kappa_j;\omega^2 r^2 \right)
\end{array}\right]\,,~~~~
\end{eqnarray}  
defined up to the normalization factors $N_1$ and $N_2$.

The difficult task now is to derive these factors according to the condition (\ref{norad}) since we have not yet general rules for normalizing the mode functions expressed in terms of hypergeometric ones in the case of the continuous energy spectra.  Nevertheless, here we can derive these quantities by using Eq. of Ref. \cite{GR} as an integral rep. as we show in  Appendix D. Thus we obtain that the normalization condition (\ref{norad}) is accomplished if we take
\begin{eqnarray}
N_1&=&\left[\sqrt{2\omega \cosh\pi\mu}\,\Gamma\left(\frac{1}{2}-\kappa_j\right)\right]^{-1}\left| \frac{\Gamma(\beta)}{\Gamma(\alpha)}\right|\,, \\
N2&=&\left[\sqrt{2\omega \cosh\pi\mu}\,\Gamma\left(\frac{1}{2}+\kappa_j\right)\right]^{-1}\left| \frac{\Gamma(\alpha)}{\Gamma(\beta)}\right|\,.
\end{eqnarray}  
Unfortunately, we cannot perform other integrals for investigating, for example, if the particular solutions are orthogonal or for calculating transition coefficients.

We remain thus with these results allowing us to write down the general fundamental spinors of this rep., $\{H_S,K,J_3\}$ selecting only the solutions regular in $r=0$ as
\begin{equation}
{\cal R}_{E,\kappa_j}=\frac{1-{\rm sign}\,\kappa_j}{2}\,{\cal R}_{E,\kappa_j}^{1} +\frac{1+{\rm sign}\,\kappa_j}{2}\, {\cal R}_{E,\kappa_j}^{2}\,.
\end{equation}
Then, in order to use a compact notation, we introduce the matrix $\Phi_{\kappa_j,m_j}=\left[  \Phi_{\kappa_j,m_j}^+, \Phi_{\kappa_j,m_j}^-\right]^T$ helping us to write down the normalized particular solutions of positive frequencies simply as
\begin{equation}
U^{S}_{E,\kappa_j, m_j}(t,r,\theta,\phi)=\frac{e^{-iEt}}{r}{\cal R}_{E,\kappa_j}(r)\Phi_{\kappa_j,m_j}(\theta,\phi)
\end{equation}
while the negative frequencies ones have to be derived by using the charge conjugation. 

Finally, we transform this basis in the equivalent basis of the rep. $\{H,K,J_3\}$ of NP as
\begin{eqnarray}
U_{E,\kappa_j, m_j}(t,r,\theta,\phi)&=&T(t)^{-1}U^{S}_{E,\kappa_j, m_j}(t,r,\theta,\phi)=U^{S}_{E,\kappa_j, m_j}(t,r e^{\omega t},\theta,\phi)\nonumber\\
&=&\frac{e^{-(\omega+iE)t}}{r}{\cal R}_{E,\kappa_j}(e^{\omega t}r)\Phi_{\kappa_j,m_j}(\theta,\phi)
\end{eqnarray}
and similarly for antiparticle spinors. We obtain again common eigenspinors having no separated variables expected to comply with a special time evolution. 

\section{Concluding remarks}

We tried to present here exhaustively all the analytical solutions of the free Dirac field minimally coupled to the gravity of the spatially flat FLRW space-times. In addition, we report a new solution which completes our collections of pairs of plane (pl.) and spherical (sph.) waves. The next table resumes all the types of solutions discussed here

\begin{center}
\begin{tabular}{lcccl}
Basis&Rep.&Manifold&Picture&Refs.\\
&&&&\\
plane P-w. &$\{P^i,S_3\}$& FLRW&NP&\cite{nach,CD1}\\
spherical  P-w.&$\{{\bf P}^2,K,J_3\}$& FLRW&NP&\cite{SH,CD2}\\
plane. E-w.&$\{H_S,N^i,S_3\}$& dS&SP&\cite{CD4}\\
spherical E-w.&$\{H_S,K,J_3\}$&dS&SP&Sec. 5.3\\
plane E-w.&$\{H,N^i,S_3\}$& dS&NP&\cite{CD4}\\
spherical E-w.&$\{H,K,J_3\}$&dS&NP&Sec. 5.3
\end{tabular}
\end{center}

\noindent We presented first the general theory of the plane and spherical P-waves in NP of the FLRW space-time reducing the problem of finding solutions to the simple systems of equations which yield the t.m.f. governing the time evolution of the Dirac field. These may be solved in many concrete cases but here we restricted ourselves to present two examples,  i.e. the Milne type universe and the dS expanding universe. This last manifold where the energy is conserved is the only FLRW space-time laying out all the solutions listed above.

Technically speaking, we presented the framework in which the gauge covariant Dirac field can be studied on FLRW space-times. Moreover, we considered the time evolution pictures allowing us to derive the P-waves in NP and the dS E-waves in SP where the variables can be separated. Turning back in NP with the E-waves derived in SP we obtain special eigenspinors of the energy operator whose variables are no longer separated.    

All the solutions presented here are determined by sets of commuting operators up to an integration constant that must be fixed according to supplemental criteria.   For the P-waves this means to fix the vacuum by choosing between the traditional a,v. and the new r.f.v. However, for the E-waves whose frequencies are separated by construction these vacua are helpless such that we must look for alternative criteria. This problem remains open. 

We must specify that apart the new solutions of section 5 we present here for the first time the definitions of the energy and Hamiltonian operators in FLRW space-times, the Minkowskian projection, the most general form of the P-waves, the closed form of the plane E-waves as well as the transition coefficients between the bases of plane P-waves and E-waves derived in section 5.2.

Finally, we hope the results presented here will open new possibilities of integrating the Dirac field in a large QFT on FLRW space-times with applications in astrophysics and cosmology.

\appendix

\section{Pauli spinors}
\setcounter{equation}{0} \renewcommand{\theequation}
{A.\arabic{equation}}

Given an arbitrary direction of unit vector ${\bf n}$, the Pauli spinors $\xi_{\sigma}({\bf n})$ defined as
\begin{eqnarray}
\xi_{\frac{1}{2}}({\bf n})&=&\sqrt{\frac{1+n^3}{2}}\left(
\begin{array}{c}
1\\
\frac{n^1+i n^2}{1+n^3}
\end{array}\right)\,,\label{spin1}\\ 
\xi_{-\frac{1}{2}}({\bf n})&=&\sqrt{\frac{1+n^3}{2}}\left(
\begin{array}{c}
\frac{-n^1+i n^2}{1+n^3}\\1
\end{array}\right)\,,\label{spin2}
\end{eqnarray}
and the conjugated ones, $\eta_{\sigma}({\bf n})=i\sigma_2 \xi_{\sigma}({\bf n})^*$.
form an arbitrary spin basis  satisfying the eigenvalue equations
\begin{eqnarray}
&&({\bf n}\cdot {\bf \sigma})\,\xi_{\sigma}({\bf n})=2\sigma \xi_{\sigma}({\bf n})\,,\\
&& ({\bf n}\cdot {\bf \sigma})\,\eta_{\sigma}({\bf n})
=-2\sigma\eta_{\sigma}({\bf n})\,,
\end{eqnarray} 
where the polarization $\sigma=\pm\frac{1}{2}$ gives the projection of SPin on the direction ${\bf n}$.
In the Dirac theory the direction ${\bf n}$ is defined in the rest frame where ${\bf p}=0$.
In current applications one takes ${\bf n}={\bf e}_3$ along the third axis of this frame.  
We must specify that in the momentum-spin rep.  we do not have a corresponding differential operator since SPin projection is defined in the rest frames.  

Another choice is the helicity basis where ${\bf n}={\bf n}_p$ is along the momentum direction. In this basis the spinors $\xi_{\lambda}({\bf n}_p)$ and  $\eta_{\lambda}({\bf n}_p)=i\sigma_2 \xi_{\lambda}({\bf n}_p)^*$ depend on the helicity $\lambda=\pm \frac{1}{2}$ which is proportional to the eigenvalues of the Pauli-Lubanski operator (\ref{PaLu}). 

\section{Spherical Dirac spinors}
\setcounter{equation}{0} \renewcommand{\theequation}
{B.\arabic{equation}}

The Dirac spherical spinors, $\Phi^{\pm}_{\kappa_j,m_j}: S^2 \to {\Bbb C}$, solve  the eigenvalue problems of the commuting operators $\{{\bf J}^2,J_3,K\}$  for the eigenvalues $\{j(j+1),m_j,-\kappa_j\}$ which can take the values $j=\frac{1}{2},\frac{3}{2},...$, $m_j=-j,-j+1,...,j$ and $\kappa_j=\pm\left(j+\frac{1}{2}\right)$. The operator eigenvalues $l(l+1)$ of the operator ${\bf L}^2$ give the orbital angular quantum number $l$ such that $j=l\pm\frac{1}{2}$ \cite{TH}. The quantum numbers $l$ and $j$ do not appear explicitly since, 
\begin{equation}\label{kjl}
\kappa_j=\left\{\begin{array}{lcc}
~~~~\,j+\frac{1}{2}=l&{\rm for}& j=l-\frac{1}{2}\\
-(j+\frac{1}{2})=-l-1&{\rm for}& j=l+\frac{1}{2}
\end{array}\right.
\end{equation}
encapsulate all of them,  $j=|\kappa_j|-\frac{1}{2}$ and $l=|\kappa_j|-\frac{1}{2}(1-{\rm sign}\, \kappa_j)$ \cite{TH,LL}.  

The above angular spinors are expressed in terms of the well-known Pauli spherical spinors, $\Psi^{m_j}_{j\pm\frac{1}{2}}$, as
\begin{equation}
 \Phi^{+}_{\mp(j+\frac{1}{2}),m_j}=\left(
 \begin{array}{c}
 i\Psi^{m_j}_{j\mp\frac{1}{2}}\\
 0
 \end{array}\right)\,, \quad
  \Phi^{-}_{\mp(j+\frac{1}{2}),m_j}=\left(
 \begin{array}{c}
 0\\
 \Psi^{m_j}_{j\pm\frac{1}{2}}\\
 \end{array}\right)\,,
\end{equation}
forming  an orthonormal set,
\begin{eqnarray}
\left<\Phi^{\pm}_{\kappa_j,m_j}, \Phi^{\pm}_{\kappa'_j,m'_j}\right>&=&\delta_{\kappa_j,\kappa_j'}\delta_{m_j,m_j'}\,,\\
\left<\Phi^{\pm}_{\kappa_j,m_j}, \Phi^{\mp}_{\kappa'_j,m'_j}\right>&=&0\,,
\end{eqnarray}
with respect to the angular scalar product
\begin{equation}\label{spSA}
\left< \Phi, \Phi'\right>=\int_{S^2}d\Omega\, \Phi(\theta,\phi)^*\Phi'(\theta,\phi)
\end{equation} 
defined on the sphere $S^2$. 

the spherical spinors help us to separate the spherical variables $(r, \theta,\phi)$ associated to  ${\bf x}$ (with $r=|{\bf x}|$) by using the following identities \cite{TH},
\begin{equation}\label{identS}
\frac{{\bf \sigma}\cdot{\bf x}}{r}\Psi_{j\pm\frac{1}{2}}^{m_j}=\Psi_{j\mp\frac{1}{2}}^{m_j} ~\to~i\frac{{\bf \gamma}\cdot{\bf x}}{r}\Phi_{\kappa_j,m_j}^{\pm}=\Phi_{\kappa_j, m_j}^{\mp}\,,
\end{equation}
and observing that $\gamma^0 \Phi_{\kappa_j,m_j}^{\pm}=\pm \Phi_{\kappa_j,m_j}^{\pm}$.

For performing the charge conjugation we take into account that the spherical harmonics satisfy $(Y_l^m)^*=(-1)^mY_{-m}^l$ such that we can write 
\begin{equation}
i\sigma_2\left(\Psi^{m_j}_{j\pm\frac{1}{2}}\right)^*=\mp (-1)^{m_j+\frac{1}{2}}\Psi^{-m_j}_{j\pm\frac{1}{2}}\,.
\end{equation}
from which we deduce:
\begin{equation}\label{chsph}
i\gamma^2\left(\Phi^{\pm}_{m_j,\mp(j+\frac{1}{2})} \right)^*=\pm(-1)^{m_j}\Phi^{\mp}_{-m_j, \pm(j+\frac{1}{2})}\,.
\end{equation}

\section{Some properties of Bessel functions}
\setcounter{equation}{0} \renewcommand{\theequation}
{C.\arabic{equation}}

The Bessel functions $J_{\nu}(z)$ and $J_{-\nu}(z)$ form a satisfactory set of independent solutions whose Wronskian gives the identity  \cite{NIST}
\begin{equation}\label{JJJJ}
J_{\nu+1}(z)J_{-\nu}(z)+J_{-\nu-1}(z)J_{\nu}(z)=-\frac{2}{\pi z}\sin\pi\nu\,.
\end{equation}
With their help one defines the modified Bessel functions 
\begin{eqnarray}
I_{\nu}(z)&=&e^{\pm \frac{i\pi\nu}{2}} J_{\nu}\left(z e^{\pm\frac{i\pi}{2}}\right)\,,\\
K_{\nu}(z)&=&K_{-\nu}(z)=\frac{\pi}{2}\,\frac{I_{-\nu}(z)-I_{\nu}(z)}{\sin\pi\nu}\,.\label{KII}
\end{eqnarray}
The functions  $K_{\nu_{\pm}}(z)$, with
$\nu_{\pm}=\frac{1}{2}\pm i \mu$ are related among themselves through
\begin{equation}\label{H1}
[K_{\nu_{\pm}}(z)]^{*}
=K_{\nu_{\mp}}(z^*)\,,\quad \forall z \in{\Bbb C}\,,
\end{equation}
satisfying the equations
\begin{equation}\label{H2}
\left(\frac{d}{dz}+\frac{\nu_{\pm}}{z}\right)K_{\nu_{\pm}}(z)=-K_{\nu_{\mp}}(z)\,,
\end{equation}
and the identities
\begin{equation}\label{H3}
K_{\nu_{\pm}}(i x)K_{\nu_{\mp}}(-i x)+ K_{\nu_{\pm}}(-i x)K_{\nu_{\mp}}(i x)=\frac{\pi}{ |x|}\,,
\end{equation}
that guarantees the correct orthonormalization properties of the fundamental spinors. 

For $|z|\to \infty$  we have  \cite{NIST}
\begin{equation}\label{Km0}
I_{\nu}(z) \to \sqrt{\frac{\pi}{2z}}e^{z}\,, \quad K_{\nu}(z) \to K_{\frac{1}{2}}(z)=\sqrt{\frac{\pi}{2z}}e^{-z}\,,
\end{equation} 
for any $\nu$, while for $z\to 0$ these functions behave as
\begin{equation}\label{beh}
I_{\nu}(z)\sim \frac{1}{\Gamma(\nu+1)}\left(\frac{z}{2}\right)^{\nu}\,, \quad  K_{\nu}(z)\sim \frac{1}{2}\Gamma(\nu)\left(\frac{z}{2}\right)^{-\nu}\,.
\end{equation}

\section{Normalization integrals}
\setcounter{equation}{0} \renewcommand{\theequation}
{D.\arabic{equation}}

In spherical coordinates of the momentum space, ${\bf n}_p\sim
(\theta_n,\phi_n)$, and the notation ${\bf p}=\omega s{\bf n}_p$, we have $d^3p=
p^2dp\, d\Omega_n=\omega^3\, s^2ds\, d\Omega_n$ with
$d\Omega_n=d(\cos\theta_n)d\phi_n$. Moreover, we can write
\begin{eqnarray}
\delta^3({\bf p}-{\bf p}^{\,\prime})&=&\frac{1}{p^2}\,\delta(p-p')\delta^2({\bf n}_p-{\bf n}_p') \nonumber\\
&=&\frac{1}{\omega^3 s^2}\,\delta(s-s')\delta^2({\bf n}_p-{\bf n}_p')\,,\label{del}
\end{eqnarray}
where  $\delta^2({\bf n}_p-{\bf n}_p') =\delta(\cos \theta_n-\cos
\theta'_n)\delta(\phi_n-\phi'_n)\,.$
Then the scalar products of the fundamental spinors of positive frequencies can
be calculated according to Eqs. (\ref{US12}), (\ref{fiplus1}), (\ref{fiplus2})  and (\ref{del}) as
\begin{eqnarray}
&&\left<U^{S\, a}_{E,{\bf n},\sigma},U^{S\, b}_{E,{\bf n}^{\,\prime},\sigma^{\prime}}\right>_S= \int_D d^3x\, [{U}^{S\,a}_{E,{\bf
n},\sigma}(t,{\bf x})]^+ U^{S\,b}_{E,{\bf n}^{\,\prime},\sigma^{\prime}}(t,{\bf
x})
\nonumber\\
&&~~~~=\delta_{ab}\frac{(2\pi)^4}{\omega^2}|N_a|^2 e^{i(E-E')t}\left[\frac{1}{2\pi\omega}\int_0^{\infty}\frac{ds}{s}\,
e^{i(\epsilon-\epsilon')\ln s}\right] \delta_{\sigma\sigma^{\prime}}\,
\delta^2
({\bf n}-{\bf n}^{\,\prime})\nonumber\\
&&~~~~=\delta_{ab}\frac{(2\pi)^4}{\omega^2}|N_a|^2\delta(E-E')\,\delta_{\sigma\sigma^{\prime}}\, \delta^2({\bf n}-{\bf n}^{\,\prime})\,,~~~~~a,b=1,2\,,~~~~\label{IntUU}
\end{eqnarray}
where we used the following rep.  of the Dirac $\delta$ -function
\begin{equation}\label{XD}
\delta(E-E')=\frac{1}{2\pi\omega}\int_0^{\infty}\frac{ds}{s}\,
e^{i(\epsilon-\epsilon')\ln s}\,.
\end{equation}
Eqs.  (\ref{orto1}) and (\ref{orto2}) are deduced in the same manner.

For calculating the normalization condition (\ref{norad}) we use Eq.  (6.574) of Ref. \cite{GR} as an integral representation,
\begin{eqnarray}
F(a,b;c;x^2)&=&x^{1-c}\frac{2^{c-b-a}\Gamma(c)\Gamma(1-b)}{\Gamma(a)}\nonumber\\
&\times&\int_{0}^{\infty}ds\, s^{a+b-c}J_{c-1}(xs)J_{a-b}(s)\,, \label{Intrep}
\end{eqnarray}
and perform first the radial integral  according to Eqs. (\ref{radnorm}) and (\ref{JJJJ}) obtaining an intermediate result proportional to $\delta(s-s')$. Furthermore, we integrate over $s'$ remaining with an integral over $s$ giving just the Dirac $\delta$-function as in Eq. (\ref{IntUU}).


\begin{thebibliography}{20}

\bibitem{nach} 
O. Nachtmann, {\it Comm. Math. Phys.} {\bf 6} (1967) 1.

\bibitem{OT}
V. S. Otchik, {\em Class. Quant. Grav.}{\bf 2}  (1985)  539.

\bibitem{BADU}
A. O. Barut and I. H. Duru, {\em Phys. Rev. D} {\bf 36}  (1987) 3705.

\bibitem{ALG}
I. E. Andrushkevich and G. V. Shishkin, {\em Theor. Math. Phys.} {\bf 70} (1987) 204. 

\bibitem{SH}
G. V. Shishkin and V. M. Villalba, {\em J. Math. Phys.} {\bf 30}  (1989)  2132.

\bibitem{Vil}
V. M. Villalba and U. Percoco,  {\em J. Mafh. Phys.} {\bf 31}  (1989) 715.

\bibitem{SH1}
G. V. Shishkin, {\em Class. Quantum Grav.} {\bf 8} (1991)  175.

\bibitem{FGV}
F. Finelli, A. Gruppuso and G. Venturi, {\em Class. Quantum Grav.} {\bf 16}  (1999) 3923.

\bibitem{KT}
V. V. Klishevich and V. A. Tyumentsev, {\em Class. Quantum Grav.} {\bf 22}  (2005) 4263.

\bibitem{CD1}
I. I. Cot\u aescu, {\em Phys. Rev. D} {\bf 65}  (2002)  084008.

\bibitem{CD2}
I. I.  Cotaescu, R. Racoceanu and C. Crucean {\em Mod. Phys. Lett. A} {\bf 21}  (2006) 1313.

\bibitem{CD3}
I. I. Cotaescu,  {\em Mod. Phys. Lett. A} {\bf  22}  (2007) 2965.

\bibitem{CD4}
I. I. Cotaescu  and C. Crucean 
Int. J. Mod. Phys. A    23 , 3707-3720 (2008)

\bibitem{CD5}
I. I. Cot\u aescu, {\em Mod. Phys. Lett. A}  {\bf 22}  (2011) 1613.

\bibitem{CD6}
I. I. Cot\u aescu and D.M. Baltateanu, {\em Mod. Phy. Lett. A} {\bf 30}  (2015) 1550208.

\bibitem{CD7}
I. I Cot\u aescu, {\em Int. J. Mod. Phys. A} {\bf 33}  (2018) 1830007.

\bibitem{dSP1}
R. Gautreau, {\em Phys. Rev. D} {\bf 27} (1983) 764.

\bibitem{dSP2}
M. K. Parikh, {\em Phys. Lett.} {\bf B 546} (2002) 189.

\bibitem{BuD}
T. S. Bunch and P. C. W. Davies, {\em Proc. R. Soc. London} {\bf 360}, 117 (1978).

\bibitem{BD}
N. D. Birrell and P. C. W. Davies,  {\em Quantum Fields in Curved
Space} (Cambridge University Press, Cambridge 1982).

\bibitem{CrfvD}
I. I. Cot\u aescu, {\em Eur. Phys. J. C} {\bf 79} (2019) 696.

\bibitem{ES}
I. I. Cot\u aescu {\em J. Phys. A: Math. Gen.} {\bf 33}  (2000) 9177.

\bibitem{dS}
I. I. Cot\u aescu, {\em Mod.  Phys. Lett. A}  {\bf 28}  (2013) 1350033.

\bibitem{CML}
B. Carter  and  R. G. McLenaghan  {\em Phys. Rev. D} {\bf 19} (1979) 1093.

\bibitem{CCML}
I. I. Cot\u aescu, {\em Europhys. Lett.} {\bf 86}  (2009)  20003.

\bibitem{CGRG}
I. I. Cot\u{a}escu, {\em GRG} {\bf 43} (2011) 1639.

\bibitem{BDR}
S. Drell and J. D. Bjorken, {\em Relativistic Quantum Fields} (McGraw-Hill Book Co., New York 1965).

\bibitem{CQED}
I. I. Cot\u aescu and C. Crucean, {\em Phys. Rev. D} {\bf 87} (2013) 044016.


\bibitem{LL}
V. Berestetski, E. Lifchitz, L. Pitayevski, {\em Th\' eorie
Quantique Relativiste} (Mir, Moscow, 1972).

\bibitem{Milne}
I. I. Cot\u aescu and D. Popescu,  {\em Chinese Phys. C}  {\bf 44} (2020) 055104.

\bibitem{prep}
I. I. Cot\u aescu, arXiv:2003.00444. 

\bibitem{KG}
I. I. Cotaescu , C. Crucean and A. Pop,  {\em Int. J. Mod. Phys. A} {\bf  23}  (2008) 2563. 

\bibitem{Max}
I. I. Cot\u aescu and C. Crucean, {\em Prog. Theor. Phys.} {\bf 124} (2010) 1051.  

\bibitem{TH}
B. Thaller,  {\em The Dirac Equation} (Springer Verlag, Berlin
Heidelberg, 1992).


\bibitem{NIST}
F. W. J. Olver, D. W. Lozier, R. F. Boisvert and C. W. Clark, {\em NIST Handbook of Mathematical Functions} (Cambridge University Press, 2010).

\bibitem{GR}
I. S. Gradshtein and I. M. Ryzhik, {\em Table of Integrals,
Series, and Products} (Academic Press Inc., San Diego, 1980).




\end{thebibliography}
\end{document}